%% file: conf_sum2004.tex
\documentclass[11pt]{article}
\usepackage{graphicx}
\usepackage{setspace}

\newcommand{\BABARPubYear}    {04}

\newcommand{\BABARConfNumber} {043}
\newcommand{\SLACPubNumber} {10637}

\input babarsym
\def\rb{\ensuremath{{r_\B}}}
\def\rbs{\ensuremath{{r^*_\B}}}
\long\def\inst#1{\par\nobreak\kern 4pt\nobreak
    {\it #1}\par\vskip 10pt plus 3pt minus 3pt}

\begin{document}
{\pagestyle{empty}

\begin{flushright}
\babar-CONF-\BABARPubYear/\BABARConfNumber \\
SLAC-PUB-\SLACPubNumber \\
December 2004 \\
\end{flushright}

\par\vskip 5cm

\begin{center}
\Large \bf 
 Constraints on \rb~ and $\gamma$ in $B^\pm\rightarrow D^{(\ast)0}K^\pm$ decays by a Dalitz analysis of $D^0 \rightarrow   K_S \pi^- \pi^+$
\end{center}
\bigskip

\begin{center}
\large The \babar\ Collaboration\\
\mbox{ }\\
\today
\end{center}
\bigskip \bigskip

\begin{center}

\input{abstract-prl}

\end{center}

\vfill
\begin{center}

Submitted to the 32$^{\rm nd}$ International Conference on High-Energy Physics, ICHEP 04,\\
16 August---22 August 2004, Beijing, China

\end{center}

\vspace{1.0cm}
\begin{center}
{\em Stanford Linear Accelerator Center, Stanford University, 
Stanford, CA 94309} \\ \vspace{0.1cm}\hrule\vspace{0.1cm}
Work supported in part by Department of Energy contract DE-AC03-76SF00515.
\end{center}

\newpage
} 

%
%
\input authors_sum2004.tex

\section{INTRODUCTION}
\label{sec:Introduction}
\input{intro-prl}

\section{THE \babar\ DETECTOR AND DATASET}
\label{sec:babar}
\input{detector-prl}

\section{EVENT SELECTION}
\label{sec:eventselection}
\input{selection-prl}

\section{DETERMINATION  OF $D^0 \rightarrow K_S \pi^- \pi^+$  DECAY MODEL}
\label{sec:dalitzModel}
\input{dalitz-prl}

\section{CP FIT TO $B \rightarrow D^{(\ast)0} K$  SAMPLES}
\label{sec:analysis}

\input{fit-prl}

\input{results-prl}

\input{constraints}

\section{SYSTEMATIC UNCERTAINTIES}
\label{sec:Systematics}

\input{systematics-prl}

\section{RESULTS AND SUMMARY}
\label{sec:Summary}
\input{conclusions-prl}

\section{ACKNOWLEDGMENTS}
\label{sec:Acknowledgments}

%

\input acknowledgements

\input{biblio-prl}
\end{document}

%% file: abstract-prl.tex
\begin{abstract} 
We report on a study of direct  $CP$ violation in the decay $B^- \rightarrow 
D^{(*)0} K^-$  with a Dalitz analysis of the  $D^0 
\rightarrow   K_S \pi^- \pi^+$  decay using a sample of 227 million \BB pairs collected by the \babar\ detector. Reference to the charge-conjugate state is implied here. We constrain the amplitude ratio \rb $= \frac{| A ( B^- \ra \overline{D}^0 K^-)|}{| A( B^- \ra D^0 K^-)|}$  to be  $< \, 0.19$ at the $90\%$ confidence level  and  \rbs $= \frac{|A( B^- \ra \overline{D}^{*0} K^-)|}{| A( B^- \ra D^{*0} K^-)|} = 0.155 ^{+0.070}_{-0.077} \pm 0.040 \pm 0.020$   and we measure  the relative strong phase    $\delta_B=(114 \pm 41 \pm 8  \pm 10 )^\circ$ 
 and  $\delta^*_B=(303 \pm 34 \pm  14 \pm 10 )^\circ$ between the amplitudes  $A ( B^- \ra \overline{D}^{(*)0} K^-)$ and  $A( B^- \ra D^{(*)0} K^-)$. From these samples  we measure  $\gamma =(70 \pm 26  \pm 10 \pm 10)^\circ$. The first error is statistical, the second error accounts for experimental uncertainties and the third error reflects   the Dalitz model uncertainty. For this preliminary result we have quoted confidence intervals   obtained with a Bayesian technique assuming a uniform prior  in \rb, $\gamma$ and $\delta_B$. 
\end{abstract}

%% file: authors_sum2004.tex
\begin{center}
\small

The \babar\ Collaboration,
\bigskip

%
B.~Aubert,
R.~Barate,
D.~Boutigny,
F.~Couderc,
J.-M.~Gaillard,
A.~Hicheur,
Y.~Karyotakis,
J.~P.~Lees,
V.~Tisserand,
A.~Zghiche
\inst{Laboratoire de Physique des Particules, F-74941 Annecy-le-Vieux, France }
% ...

\end{center}\newpage

%% file: intro-prl.tex
In the past years  $CP$ violation in the $B$ meson
system has been clearly established \cite{ref:CP} and
although there is good agreement with the expectations 
of the Standard Model, further measurements of $CP$ violation 
in $B$ decays are needed to over-constrain the unitarity triangle
and look for New Physics effects.
A crucial test will be represented by the measurement 
of $\gamma$, which is the complex phase of the Cabibbo-Kobayashi-Maskawa \cite{ref:CKM} quark mixing matrix element $V_{ub}$  in the Wolfenstein parameterization \cite{ref:Wolf}. \\
Various methods   using $B^- \rightarrow D^0K^-$  decays \cite{chargeconj} have been proposed to measure the unitarity triangle angle $\gamma$, 
all exploiting the fact that a  $B^-$ can decay into a 
$D^0 K^-$ final state via a $b \to c $ transition  or into a  $\overline{D}^0 K^-$  final state  via a $b \to u $ transition. $CP$ violation can be detected if the $D^0$ and $\overline{D}^0$ decay into the same final state. 
The measurement of  direct $CP$ violation is sensitive 
to the phase difference between $V_{ub}$ and $V_{cb}$ and thus to the angle $\gamma$. 
 Most of the experimental methods to extract $\gamma$ can be grouped in two categories:
the $D^0$ and $\overline{D}^0$ decay into a  $CP$ eigenstate \cite{ref:GLW}; or the $D^0$  decays to a common flavor state, either through a Cabibbo-allowed or a doubly Cabibbo-suppressed mode \cite{ref:ADS}. The measurement of $\gamma$ in both methods also requires  the
knowledge of \rb, the magnitude  of the ratio of the amplitudes $\cal{A}$$(B^- \ra \overline{D}^0 K^-)$ and $\cal{A}$$(B^-\ra D^0 K^-)$  and of their  
relative strong phase $\delta_B$,  which can be obtained  from  data.\\
In this paper we report on a measurement of direct $CP$ violation in $B^- \rightarrow D^{(*)0}K^-$ based on the analysis of the Dalitz distribution of    the three-body decay  $D^0 \to  K_S\pi^-\pi^+$  \cite{ref:DKDalitz}. The advantage of this method is that it involves  the entire  resonant  substructure of the three-body decay, with  Cabibbo-allowed and doubly Cabibbo-suppressed  amplitudes interfering directly. It is therefore  expected to have  a higher statistical precision than the methods outlined above. Results of an analysis based on this procedure were reported by the Belle Collaboration in \cite{Poluektov:2004mf}. From the combination  of the $B^- \ra D^0 K^-$ and  $B^- \ra D^{*0} K^-$ mode they obtain the value $\gamma \, = \, 77^{\circ}$$^{+17^{\circ}}_{-19^{\circ}} \pm 13^{\circ} \pm 11^{\circ}$ where the first error is statistical,  the second is experimental systematics and the third is model uncertainty. They also obtain a value of \rb$=0.26 ^{+0.10}_{-0.14} \pm 0.03  \pm 0.04 $ for $B^- \ra D^0 K^-$ and \rbs~$=0.20 ^{+0.19}_{-0.17} \pm 0.02  \pm 0.04 $ for $B^- \ra D^{*0} K^-$.

\subsection{Analysis outline}

The $B^-$ and $B^+$ decay amplitudes  for the  $B^- \ra  D^{(*)0}K^-$ and $D^0  \ra K_S \pi^- \pi^+$  decays  can be written assuming no CP asymmetry in $D$ decays as :
\begin{eqnarray}
\label{eq:mamplitude}
M_-(m^2_-,m^2_+) &=& |{\cal A}(B^-\ra D^0 K^-)| \, \left[  f(m^2_-,m^2_+) + r_B e^{i(\delta_B - \gamma)}~  f(m^2_+,m^2_-) \right],\\ 
M_+(m^2_-,m^2_+) &=& |{\cal A}(B^+\ra \overline{D}^0 K^+)| \,\left[  f (m^2_+,m^2_-) + r_B e^{i(\delta_B + \gamma)}~   f  (m^2_-,m^2_+)\right], 
\label{eq:mmamplitude}
\end{eqnarray}
where $m^2_-$ and $m^2_+$ are the squared invariant masses of the $K_S
\pi^-$ and $K_S \pi^+$ combinations respectively and  $ f(m^2_-,m^2_+)$ is
the amplitude of the $D^0 \rightarrow K_S \pi^- \pi^+$ decay.

 Given a known  $f$, the bi-dimensional Dalitz $(m^2_-,m^2_+)$ distributions  for $B^-$ and $B^+$    can be simultaneously  fitted   to  $|M_-(m^2_-,m^2_+)|^2 $ and $|M_+(m^2_-,m^2_+)|^2 $ respectively. A maximum likelihood technique may be used to estimate \rb, $\delta_B$, and $\gamma$. Since the measurement of $\gamma$ arises from the interference in Eq.~\ref{eq:mamplitude} and Eq.~\ref{eq:mmamplitude}, the uncertainty in the  knowledge of the complex form  of  $f(m^2_-,m^2_+)$ can lead to a systematic uncertainty. A model describing the  $D^0 \rightarrow K_S \pi^- \pi^+$ decay in terms of two-body amplitudes has been assumed in this analysis. This  model   has been characterized  using a high statistics flavor 
tagged $D^0$ sample ($D^{\ast +} \rightarrow D^0 \pi_s^{+}$ ), obtained  from $e^+ e^- \ra c \bar c$ events  as described in Section \ref{sec:dalitzModel}.

 A similar  analysis is also performed using $B^- \rightarrow D^{*0}K^-$ decays, and 
$\gamma$ is extracted along with the amplitude ratio   \rbs~ and strong phase difference $\delta^*_B$ taking into account the effective strong phase shift of $\pi$ radians between the $D^{*0} \rightarrow D^0 \pi^0$ and $D^{*0} \rightarrow D^0 \gamma$ channels \cite{ref:bondar}. By convention $\delta^*_B$ is the strong phase of  $D^{*0} \rightarrow D^0 \pi^0$ decay mode.

%% file: detector-prl.tex
 The analysis is based  on a sample of 227 million $B\bar B$ pairs collected by the \babar\ detector at the SLAC PEP-II
 $e^+ e^-$ asymmetric-energy storage ring. \babar\ is a solenoidal detector optimized for the asymmetric-energy beams 
at PEP-II and is described in ~\cite{ref:detector}. We summarize briefly
the components that are crucial to this analysis.
Charged-particle tracking is provided by a five-layer silicon
vertex tracker (SVT) and a 40-layer drift chamber (DCH). 
In addition to providing precise spatial hits for tracking, the SVT and DCH 
also measure the ionization energy loss ($dE/dx$), which is used for particle 
identification of low-momentum charged particles. At higher momenta (p $>$ 0.7 GeV$/c$) 
pions and kaons are identified by  Cherenkov radiation detected in a ring-imaging
device (DIRC). The typical separation between pions and kaons varies from 8$\sigma$ 
at 2 GeV$/c$ to 2.5$\sigma$ at 4 GeV$/c$. Neutral cluster (photon) positions and energies are 
measured with an electromagnetic calorimeter (EMC) consisting of 6580 thallium-doped CsI crystals. 
Candidate $\pi^0$ mesons are reconstructed as pairs of photons, spatially separated in the EMC, 
with an invariant mass within 3$\sigma$ of the $\pi^0$ mass. These systems are mounted 
inside a 1.5-T solenoidal super-conducting magnet. \\

%% file: selection-prl.tex
We reconstruct the decays  $B^-\rightarrow D^0 K^-$ and 
$B^-\rightarrow D^{\ast 0} K^-$ with $D^{\ast 0} \to D^0\pi^0 \, , \,  D^0\gamma$.
A larger sample of $B^-\rightarrow D^{ (*)0}\pi^-$  is also reconstructed
 and is used as a control sample to determine the Probability Density Function (PDF)  of the discriminating variables   used in the likelihood fit for $\gamma$. \\
 $D^0$ candidates  are reconstructed in the    
$K_S\pi^+\pi^-$ final state with the $K_S$  reconstructed from pairs of oppositely charged 
pions with an invariant mass within 9~MeV$/c^2$ of the nominal $K_S$ mass \cite{ref:pdg2004}.
The two pions  are constrained  to originate from the same point.    The  angle $\alpha_{K_S}$ between  the $K_S$ line of flight and its momentum is  required to satisfy the condition $\cos{\alpha_{K_S}} > $0.99.
$D^0$ candidates are selected by making all possible combinations 
 of the  $K_S$ candidate  and two oppositely  charged pions
 with an invariant mass within 12~MeV$/c^2$ of  the nominal $D^0$ mass. \\
 
The photon candidates for $D^{*0}\rightarrow D^0\gamma$ are reconstructed from clusters in the electromagnetic calorimeter
with energy greater than 30~\mev\ and consistent with a photon shower  profile.
We select \piz\ candidates from pairs of photon candidates and require
$115<m(\gamma\gamma)<150$~\mevcc and with total energy  greater 
than 70 MeV. To improve the momentum resolution, the $\pi^0$ candidates are kinematically fitted  with their mass constrained to the nominal $\pi^0$ mass. The $D^0$ candidates are  combined with a low energy  $\pi^0$ or $\gamma$.  The $D^{\ast 0}$-$D^0$ mass difference $\Delta m$ is required to be 
within 2.5 (10) MeV$/c^2$ of the nominal $\Delta m$ for $D^{*0}\rightarrow D^0\pi^0 (\gamma)$.\\
 
 A  $B^-$ candidate is obtained by combining  a $D^{(*)0}$ 
candidate  with a track (``bachelor'' track) identified as  a  kaon as described in \cite{ref:detector}. We improve the  momentum  resolution for the $D^0$ daughters by applying a kinematic mass constraint. For every $B$ candidate two standard variables are defined,  the beam-energy-substituted mass 
$\mes\equiv\sqrt{(\frac{1}{2}s+\vec{p}_0\cdot\vec{p}_{B})^2/E_0^2-p_B^2}$
and the energy difference $\DeltaE\equiv E_B^*-\frac{1}{2}\sqrt{s}$, where
the asterisk denotes the CM frame,
$s$ is the square of the total energy in the CM frame,
$p$ and $E$ are, respectively, momentum and energy, and
the subscripts $0$ and $B$ refer to \Y4S\ and $B^{\pm}$, respectively. The resolutions, evaluated  on simulated  signal events, are $2.6$~MeV$/c^2$ and $17$~MeV  for $m_{ES}$ and $\Delta E$, respectively.

\begin{figure}[!htb]
\begin{center}
\includegraphics[height=7cm]{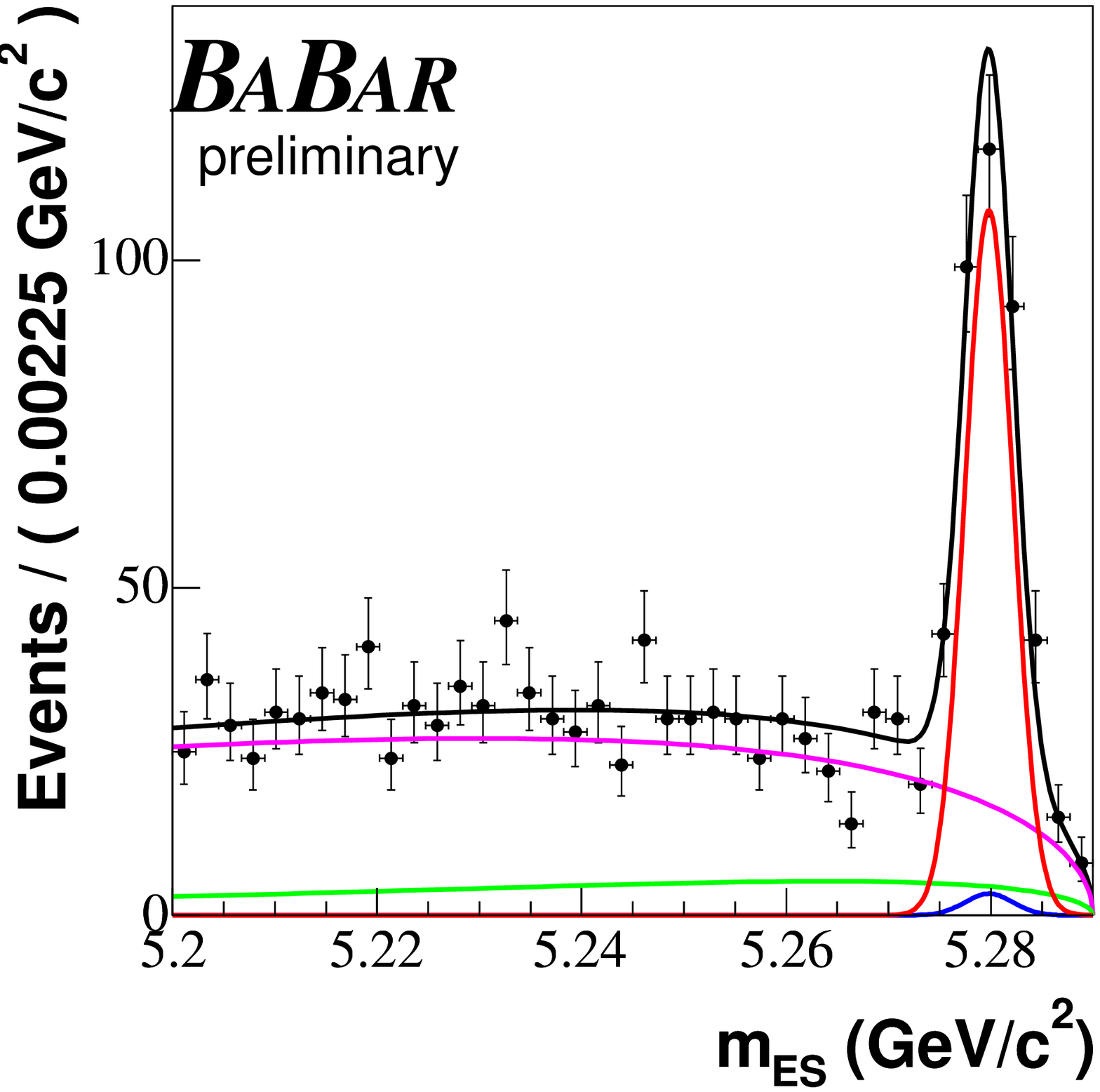}
\includegraphics[height=7cm]{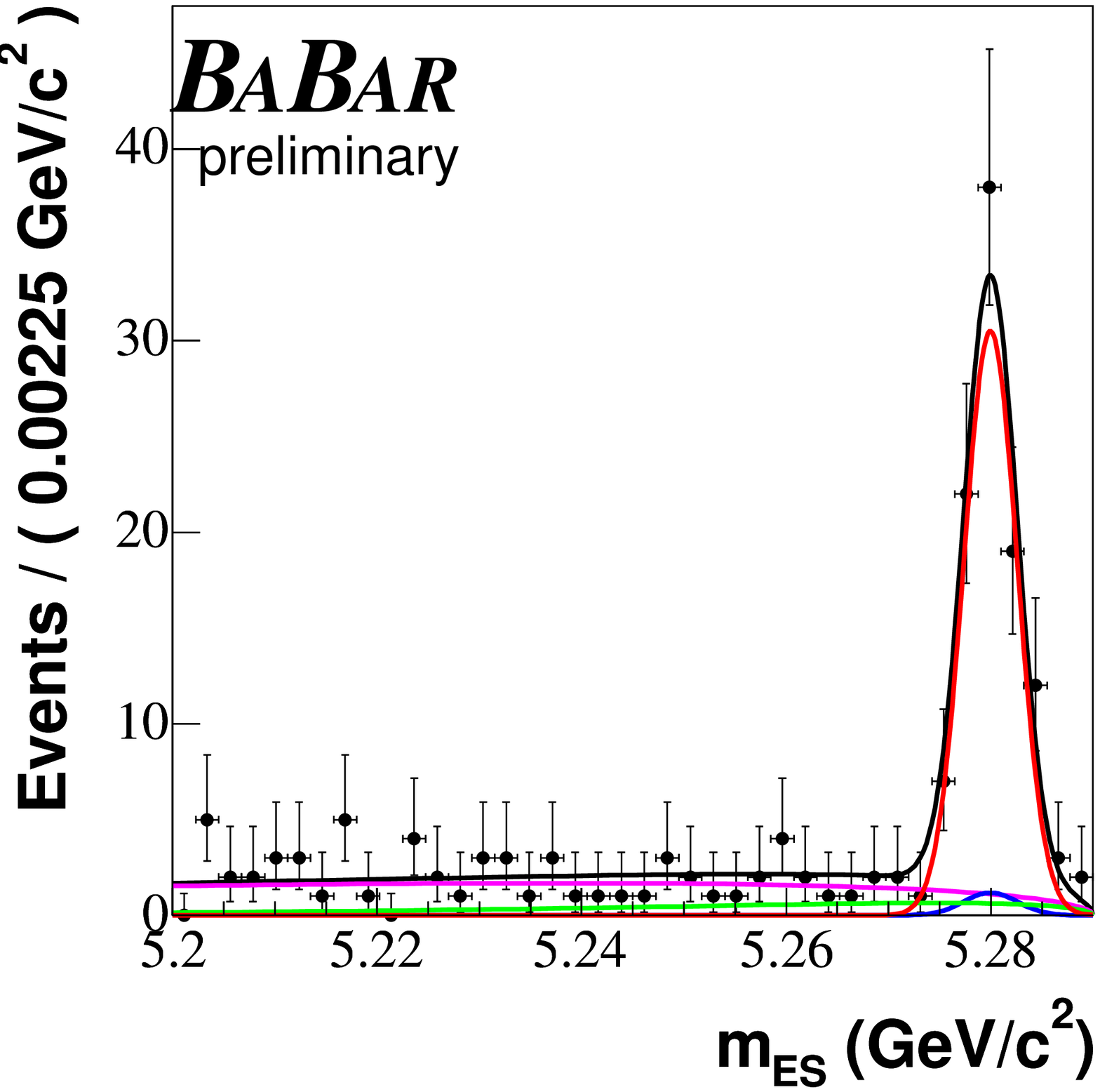}
\includegraphics[height=7cm]{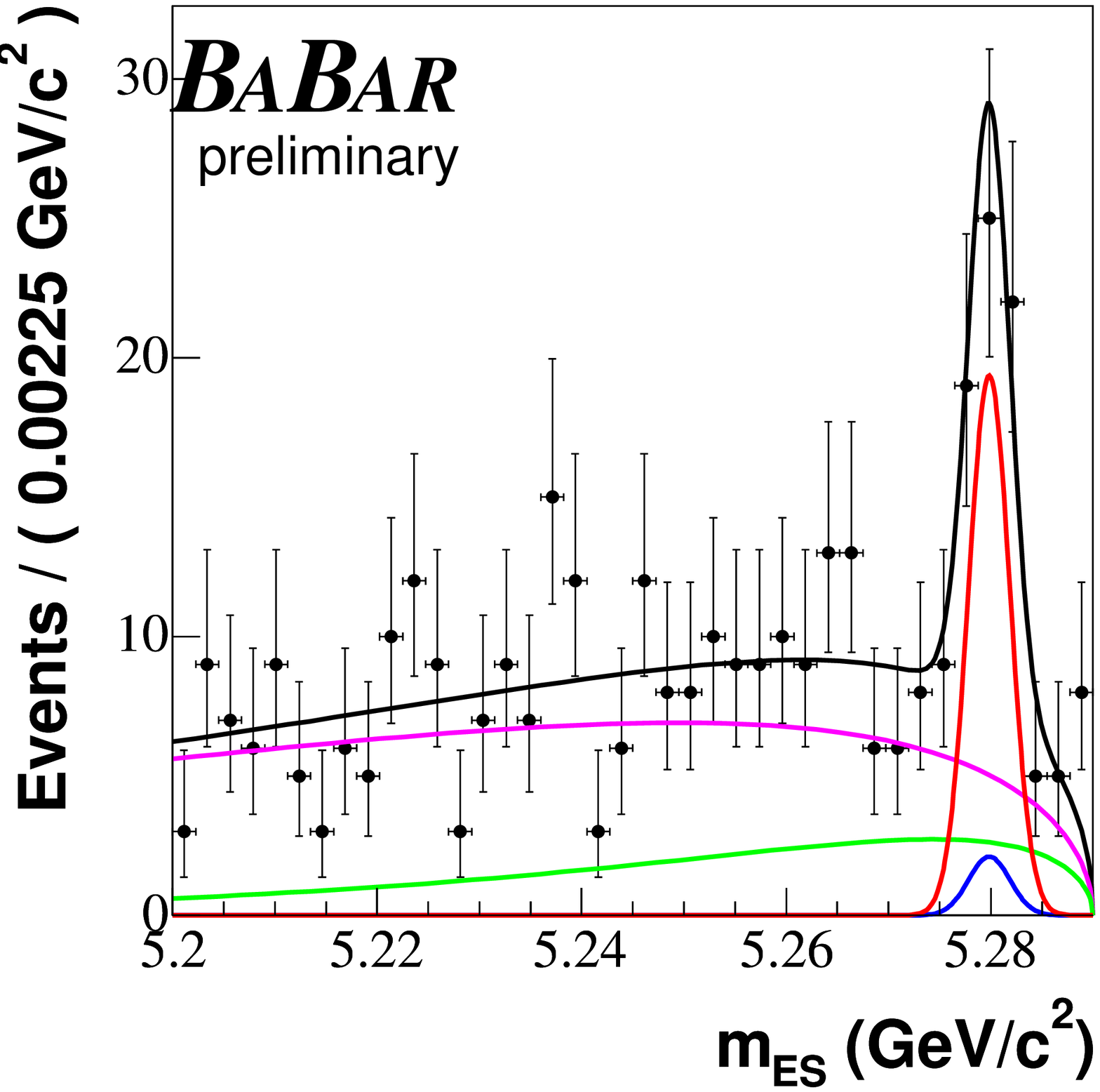}
\caption{$B^- \ra D^0 K^-$ (top left), $B^- \ra D^{*0}(D^0\pi^0) K^-$  (top right) and $B^- \ra D^{*0}(D^0\gamma)  K^-$   (bottom)  $m_{ES}$  distribution  in the $\Delta E$ region $[-30,30]$ MeV in the sample of 211 million $B\bar B$  pairs. The signal contribution is shown in red, $B^- \ra D^{*0} \pi^-$ in blue, generic $B\bar B$ in green,  and continuum in magenta.}
\label{fig:variables}
\end{center}
\end{figure}

To distinguish between $\BB$ and continuum events the following 
 topological variables  are used: $\cos \theta^\ast_B$, where $\theta^\ast_B$ 
is the polar angle of the $B$ candidate with respect to the beam axis in the CM frame; 
$L_0 = \sum_i{p_i}$ and $L_2 = \sum_i{p_i \cos^2\theta_i}$ 
calculated in the CM frame, where $p_i$ and $\theta_i$ are the momenta and  the angles 
of tracks and neutral  clusters not used to reconstruct 
the $B$ candidate with respect to its thrust axis;
 $\cos \theta_T$, where $\theta_T$ is the angle in
the CM frame between the thrust axes of the $B$ candidate and of the
rest of the event. 
 $B$ candidates are selected  by requiring $|\cos{\theta_T}|$$<$ 0.8.
 Under this condition  a Fisher discriminant $\cal F$ 
is constructed from  the variables discussed above. 
This  Fisher discriminant  is used in 
the  likelihood fit to help  distinguish  between signal and  continuum   $e^+ e^-  \to q \bar q$ ($q\,=\, u,d,s,c$)  events. The  PDF of $\cal F$   is obtained for continuum events   using  $m_{ES}$ sideband  data. The $B^- \ra D^{(*)0} \pi^-  $ events are used to obtain the $\cal F$  PDF  for $\BB$ and signal events.\\
Finally,   the $B^- \to D^{(*)0} K^-$ sample   events are required to  satisfy
   $m_{ES} > 5.2$ GeV$/c^2$,  $|\Delta E|<~$30 MeV. 
 The $B^- \ra D^{(*)0} \pi^-  $ candidates  are selected using criteria similar to  those applied for  $B^- \ra D^{(*)0} K^-$  but requiring   that the bachelor pion not  be consistent with the kaon hypothesis.  The overall reconstruction efficiencies  are $18\%,\, 4.3\%, \, 8.1\%$ for the  $D^0 K^-$, $D^{\ast 0}(D^0\pi^0)K^-$, and  $D^{\ast 0}(D^0\gamma)K^-$ decay modes, respectively.  Fig. \ref{fig:variables} shows the  $m_{ES}$ distributions after all the selection criteria are applied.

%% file: dalitz-prl.tex
\begin{figure}[!htb]
\begin{center}
\includegraphics[height=8cm]{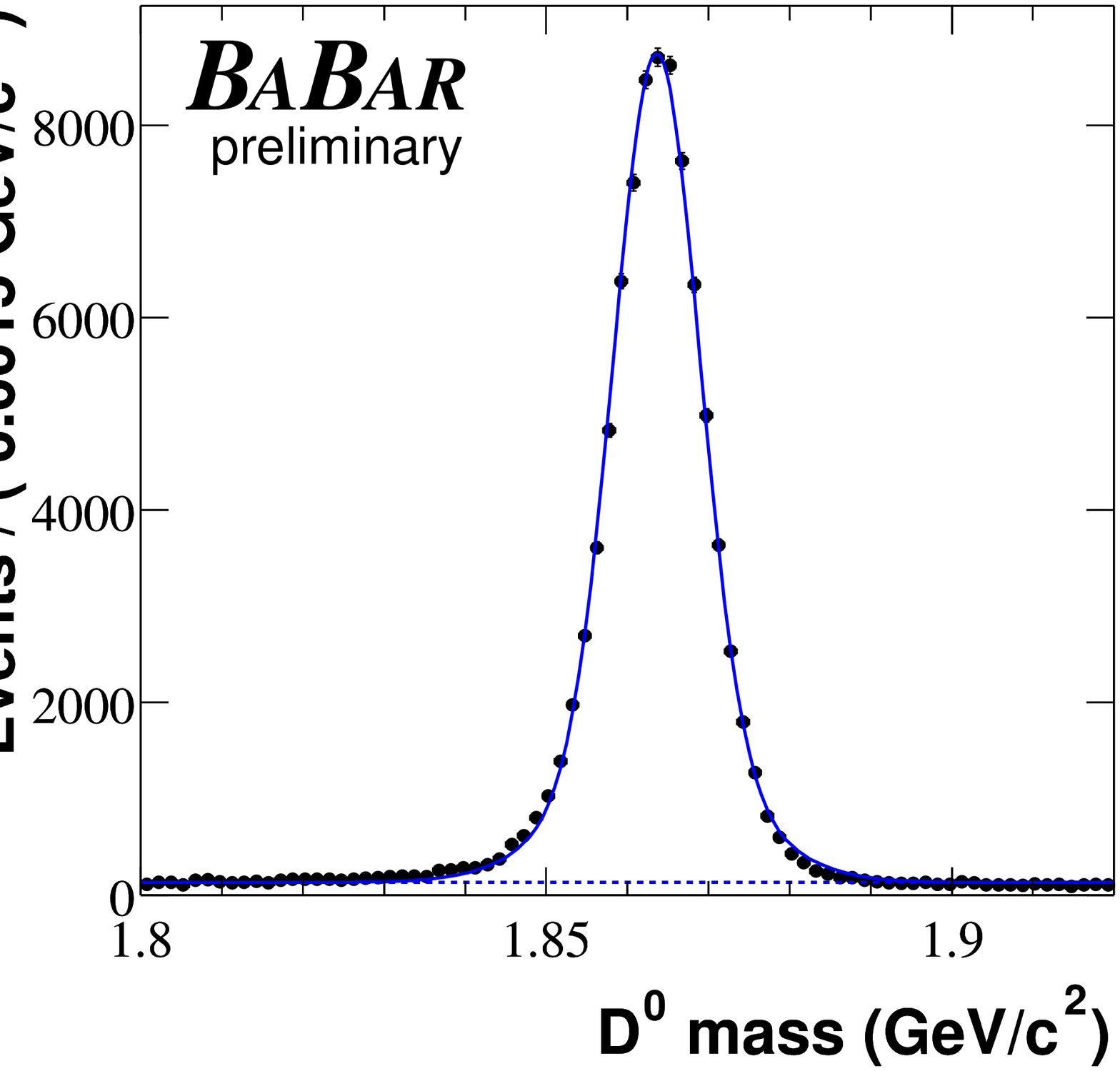}
\caption{ The $K_S \pi^- \pi^+$ invariant mass distribution  in the flavor tagged  $D^{*}$ sample. }
\label{fig:d0mass}
\end{center}
\end{figure}
\begin{figure}[!htb]
\begin{center}
\includegraphics[height=15cm]{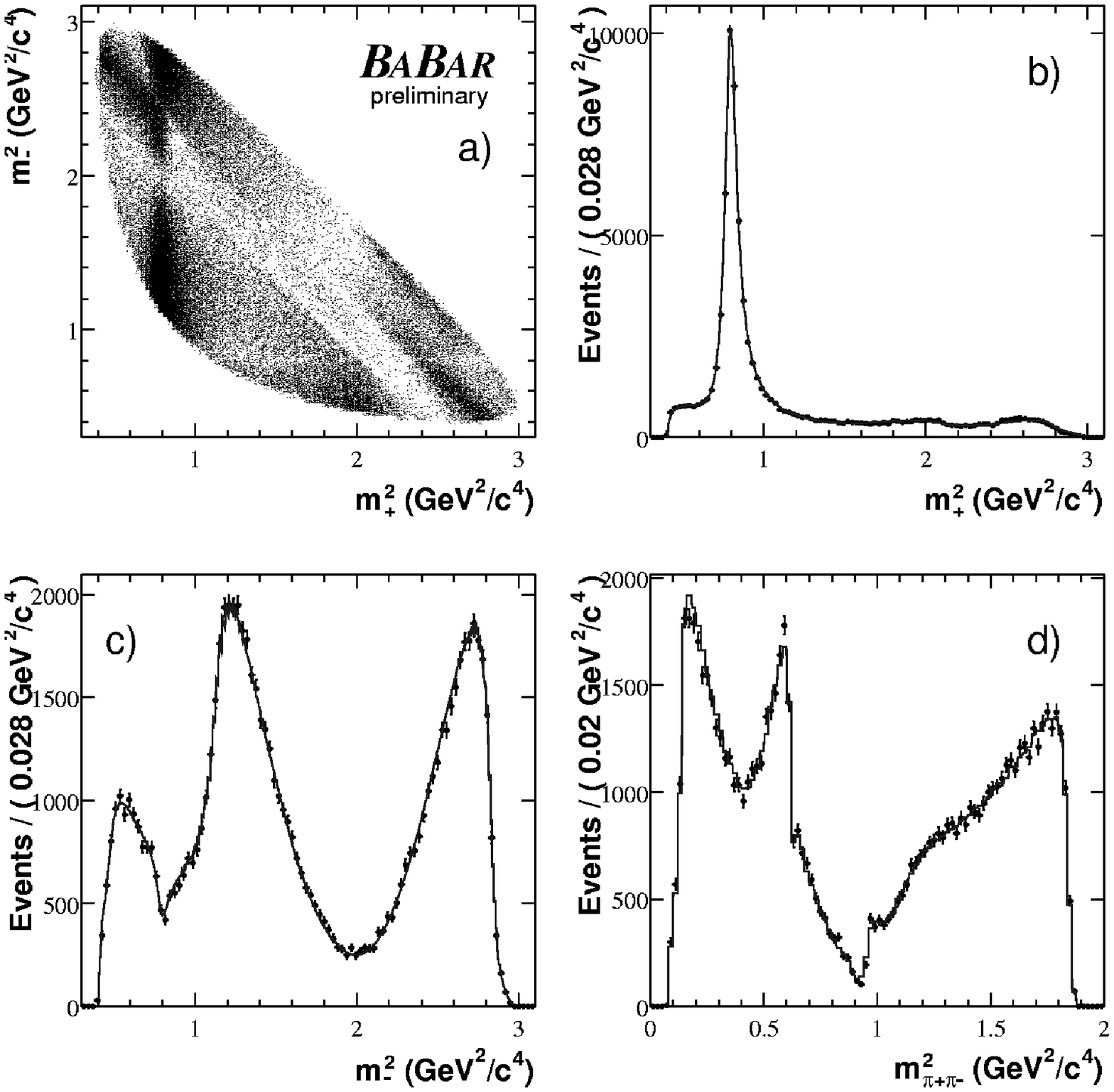}
\caption{(a) The $\overline{D}^0 \ra K_S \pi^+ \pi^-$ Dalitz distribution from the $D^{*+} \ra D^0 \pi^+$ events. Projections on (b) $m^2_+$,  (c) $m^2_-$,  and (d) $m^2(\pi^+\pi^-)$   are  shown.   The result of the fit is superimposed as a solid line. }
\label{fig:dalmkspidcs}
\end{center}
\end{figure}

 The amplitude 
$ f (m^2_\pm,m^2_\mp)$ has been constructed  from a Dalitz analysis 
 of a $97$\%  pure  flavor tagged $D^0$ sample obtained from 
  81496 $D^{*+} \rightarrow D^0 \pi^{+}$ events corresponding to a luminosity of  91.5 fb$^{-1}$ (Fig.~\ref{fig:d0mass}). 
The Dalitz $(m^2_\pm,m^2_\mp)$ distribution (Fig.~\ref{fig:dalmkspidcs}) is fitted in the context of the isobar formalism described in \cite{ref:cleomodel}. 
In this formalism the amplitude $f$ can be written as a sum of 
 two-body decay matrix elements and a non-resonant term according 
to the  expression 
\begin{equation}
  f  = a_{nr} e^{i \phi_{nr}} + 
\Sigma_r a_r e^{i \phi_r} {\cal A}_{s}(K_S \pi^- \pi^+ | r).
\end{equation}

Each term of the sum is parameterized  with an amplitude and a
phase. The factor ${\cal A}_{s}(K_S \pi^- \pi^+ | r)$ gives the Lorentz
invariant expression for the matrix element of a $D^0$ meson 
decaying into $K_S \pi^- \pi^+$ through an intermediate resonance $r$ as a
function of the  position in the Dalitz plot. It is, in general, parameterized
by a relativistic Breit-Wigner with a functional form  dependent on 
 the spin of the resonance.  For  the $\rho$   a more complex parametrization is used as  suggested in \cite{ref:gounarissakurai}.
 We fit the Dalitz distribution with a model consisting of  13 resonances  leading  to 17 two-body decay amplitudes and phases (Table \ref{tab:fitreso-likelihood}). Of the 13 resonances eight   involve a  $K_S$ plus a $\pi \pi$ 
resonance and the remaining five   are made of a  ($K_S \pi^-$)  resonance plus a $\pi^+$. We also include  the corresponding doubly Cabibbo-suppressed amplitudes  for most of the ($K_S \pi^-$) $\pi^+$ decays.  
All the  resonances considered in this model are well established except for the two scalar $\pi\pi$  resonances, $\sigma_1$ and $\sigma_2$, whose masses and widths are obtained  from our sample. Those are introduced in order to obtain an acceptable fit to the data, but their existence as true, scalar particles is a matter outside the scope of the paper. \\

An unbinned  maximum likelihood fit is performed to measure the amplitudes $a_{nr}, a_{r}$ and the phases $\phi_{nr}, \phi_r$. The fit fraction for each  decay channel is  defined as the integral of a single component divided by the coherent sum of all components. The results of the fit are shown in Fig.~\ref{fig:dalmkspidcs}.  Amplitudes, phases and fit fractions  as obtained by the  likelihood fit are  reported in Table \ref{tab:fitreso-likelihood}.
\begin{table}[htpb]

\begin{center}
\begin{tabular}{|c|c|c|c|c|c|c|}
\hline
    Resonance  &  Amplitude  & Phase  & Fraction & Mass & Width & Functional \\ 
 & & (degrees) & (\%) & MeV$/c^2$ & MeV$/c^2$ &  form \\
\hline
$K^{*}(892)$        &    $  1.777      \pm  0.018   $   &  $ 131.0     \pm  0.81  $   &   $58.51$     & 891.66       & 50.8         & BW      \\ \hline
$\rho^0(770)$       &    1 (fixed)                      &    0(fixed)                   &   $22.33$   & 775.8        & 146.4        &  GS       \\ \hline
$K^{*}(892)$ DCS    &    $  0.1789     \pm  0.0080  $   &  $-44.0        \pm  2.4 $     &    $0.59$   & 891.66      & 50.8         & BW \\ \hline
$\omega(782)$       &    $  0.0391     \pm  0.0016   $  &  $  114.8      \pm  2.5 $     &    $0.56$   & 782.6        & 8.5          & BW\\ \hline
$f_0(980) $         &    $  0.469      \pm  0.011   $  &  $   213.4     \pm  2.2   $   &    $5.81$    & 975          & 44           & BW\\ \hline
$f_0(1370) $        &    $  2.32       \pm  0.31   $    &  $   114.1     \pm  4.4   $   &    $3.39$   & 1434         & 173          & BW\\ \hline
$f_2(1270) $        &    $  0.915      \pm  0.041  $    &  $ -22.0       \pm  2.9   $   &    $2.95$   & 1275.4       & 185.1        & BW \\ \hline
$K^{*}_0(1430)$     &    $  2.454      \pm  0.074  $    &  $  -7.9      \pm  2.0   $   &    $8.37$    & 1412       & 294         & BW\\ \hline
$K^{*}_0(1430)$ DCS &    $  0.350      \pm  0.069   $   &  $  -344.       \pm  10.   $   &    $0.60$  & 1412       & 294         & BW\\ \hline
$K^{*}_2(1430)$     &    $  1.045      \pm  0.045   $   &  $  -53.1      \pm  2.6   $   &    $2.70$   & 1425.6           & 98.5  & BW\\ \hline
$K^{*}_2(1430)$ DCS &    $  0.074      \pm  0.038   $   &  $  -98        \pm  30   $    &    $0.01$   & 1425.6           & 98.5  & BW\\ \hline

$K^{*}(1410)$       &    $  0.524     \pm  0.073  $    &  $  -157       \pm  10  $     &    $0.39$    & 1414             & 232  & BW \\ \hline
$K^{*}(1680)$       &    $  0.99     \pm  0.31  $    &  $  -144       \pm  18  $     &    $0.35$      & 1717             & 322  & BW  \\ \hline
$\rho(1450)$        &    $  0.554      \pm  0.097   $   &  $  35        \pm  12.  $    &    $0.28$    & 1406             & 455  &  GS  \\ \hline
$\sigma_1$          &    $  1.346      \pm  0.044   $   &  $  -177.5      \pm  2.5  $    &    $9.11$  & $484\pm9$  & $383\pm14$ & BW \\ \hline
$\sigma_2$          &    $  0.292      \pm  0.025   $   &  $ -206.8      \pm  4.3   $   &    $0.98$  & $1014\pm7$ &  $88\pm13$  & BW\\ \hline
Non resonant        &    $  3.41      \pm  0.48  $      &  $ -233.9      \pm  5.0   $   &    $6.82$  & \-- & \-- & \-- \\ \hline
\end{tabular}
\end{center}

\caption{ Amplitudes,  phases and  fit fractions of the different components obtained from the likelihood fit of the $D^0 \rightarrow K_S \pi^-\pi^+$ Dalitz distribution in $D^{*\pm} \rightarrow D^0 \pi_s^{\pm}$ data. Masses and widths of all resonances except $\sigma_1$ and $\sigma_2$ are taken from \cite{ref:pdg2004}.  The abbreviations BW and GS  stand for relativistic  Breit-Wigner and Gounaris-Sakurai \cite{ref:gounarissakurai} respectively. The total fit fraction is $1.24$. }
\label{tab:fitreso-likelihood}
\end{table}

 We estimate the goodness of the fit for our model with a $\chi^2$ fit using adaptive  binning of the Dalitz plot. We obtain $\chi^2$/d.o.f. = 3824/(3054-32) = 1.27.

\begin{figure}[!htb]
\begin{center}
\includegraphics[height=10cm]{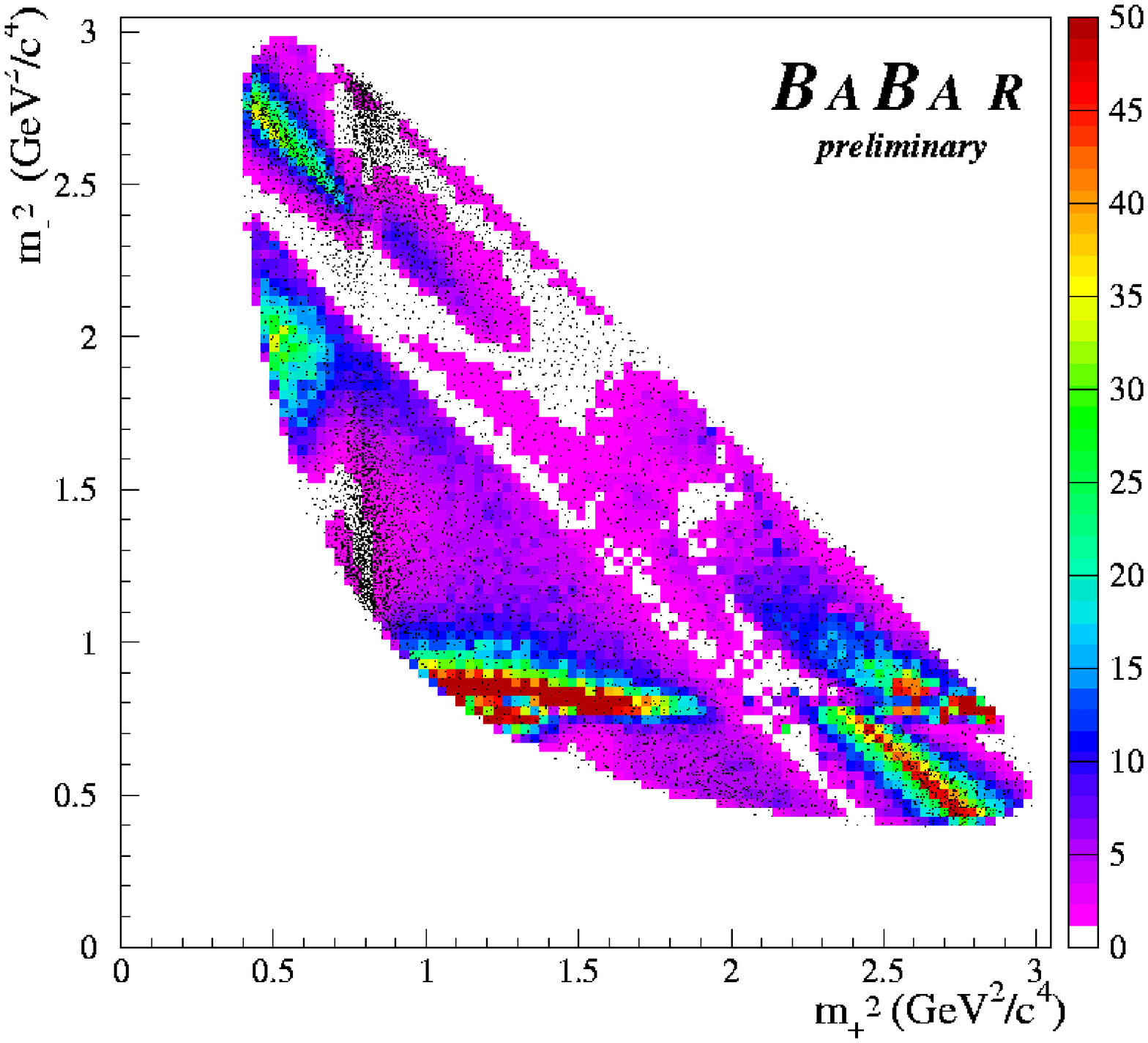}
\caption{$\overline{D}^0 \ra K_S \pi^+ \pi^-$ Dalitz distribution of simulated $ (B^+ \ra \overline{D}^0  K^+)$ events. Each event is given a weight $\frac{d^2\ln {\cal L}}{d^2\gamma}$. The black points represent the same events with weight equal to unity.}
\label{fig:gammasens}
\end{center}
\end{figure}

To illustrate the region of the Dalitz plot  most sensitive to $\gamma$ measurement,  we show in Fig.~\ref{fig:gammasens}   the distribution of simulated $ B^-\ra D^{0}  K^- $ events based on our Dalitz model, where  each event is given a weight  of  $\frac{d^2\ln {\cal L}}{d^2\gamma}$ where ${\cal L}$ is the likelihood function described in the following section.   The regions of interference between   doubly Cabibbo suppressed and Cabibbo allowed decays and $CP$ eigenstate  decays exhibit the highest sensitivity to $\gamma$.

%% file: fit-prl.tex
 A maximum likelihood fit ($CP$ fit)  is performed on the $B^- \rightarrow D^{(\ast) 0} K^-$ samples to extract simultaneously  the $CP$ violation parameters $\gamma$, $\delta_B^{(*)}$, and $r^{(*)}_B$ and the signal and background yields.  
 The likelihood for each  candidate $j$ is obtained by summing the product of the  event yield $N_{i}$ and the probability ${\cal P}_{i}$ over the  signal and the three background hypotheses.
The extended likelihood function  is 
\begin{equation}
{\cal L} = \exp{\left(-\sum_{i}N_i \right)}
\prod_{j}\left[\sum_{i}N_i{\cal P}_{i}(\vec{x}_j){\cal P}^{\rm Dal}_{i}( m^2_+, m^2_-)\right]. 
\label{eq:likelihood}
\end{equation}
The probabilities ${\cal P}_{i}$ are evaluated as the product of the PDFs 
for each of the independent variables $\vec{x}_j = \left\{\mes, \Delta E, {\cal F}\right\} $.  ${\cal P}^{\rm Dal}_{i}( m^2_+, m^2_-)$  is the PDF for the  Dalitz distribution for  the $i^{th}$ category. The categories in the fit are  signal $B^- \to D^{(*)0}K^-$, the continuum background, $B\bar B$ background, and $B^- \ra D^{*0} \pi^-$ and are shown in  Fig.~\ref{fig:variables}. 
The $m_{ES}$ and $\Delta E $ distributions for signal events  are  described  by a Gaussian. The Fisher  PDF is parametrized with a double Gaussian function. The signal PDF  parameters are determined from  the $B^- \ra D^{(*)0} \pi^-$ control sample.

 \subsection{Background Composition}

   The numbers of   events  for the various background components in  the $B^- \ra D^{(*)0}K^- $ samples are summarized in Table \ref{tab:bgfrac}.
 The dominant  background contribution  is  from the
random combination of  a real or fake $D^{(*)0}$ meson  with a charged track  in  continuum events or other $B \bar B $ decays. 
 The combinatorial background in the $m_{ES}$  distribution is described by a   threshold function \cite{ref:argus}  whose  free parameter $\xi$ is determined from the $B^- \ra D^{(\ast) 0}\pi^-$  data sample.   The shape of the combinatorial background $m_{ES}$  distribution in  generic $B \bar B $ decays   is  taken  from simulated  events.
  The  $\Delta E$ distribution   is  described by a straight line whose slope is extracted  from a fit to the  $B^- \ra D^{(\ast) 0}\pi^-$ sample. The PDF of the Fisher distribution  for continuum events  is determined from  the $m_{ES}$ sideband in the  same data sample. The  Fisher PDF for $B \bar B $ events is assumed to be the same as that for the  signal.

 An important class of background  events arises from continuum  where a  real $D^0$  is  produced back-to-back with a kaon. Depending on   the  flavor-charge correlation  this background can mimic  either the $b \to c $ or the $b \to u$   signal component. In the likelihood function we take this effect into account with two parameters,  the fraction $f_{D^0}$ of background events with a real  $D^0$  and the parameter $R$, the fraction of background events with a real  $D^0$ 
associated with an  oppositely flavored kaon (same charge correlation as the $b \to u$ signal component).

 The fraction of real  $D^0$ from continuum events  has been evaluated in events satisfying $\m_{ES}$ $<$ 5.272 GeV$/c^2$ after removing the  requirement on the $D^0$ mass. The fraction $R$ of  background events with a genuine $D^0$ associated with a negatively  charged kaon is obtained from  simulated events. The values of $f_{D^0}$ and $R$ for continuum $q \bar q$ are summarized in Table \ref{tab:bgparams}. The fraction of events with a real  $D^0$  in  generic $\BB$  events  is  found to be few percent.

 A small background originates  from   $B^- \ra D^{(*)0} \pi^-$  where the bachelor pion is misidentified  as a kaon. These events have the same   $m_{ES}$ distribution  as  signal but  can be distinguished using  their  $\DeltaE$ information.

\begin{table}[h]
\begin{center}
\begin{tabular}{|c|c|c|c|}
\hline
 Background components    &  $D^0 K^-$       & ($D^0 \pi^0$) $K^-$   & ($D^0\gamma$)$ K^-$ \\   \hline\hline
 Continuum  $q\bar q$    &  $125\pm 6$      & $9 \pm 2$                & $38 \pm 3$  \\   \hline
 $B \bar B$              &  $28 \pm 7$     & $4 \pm 2$                 & $14 \pm 5$  \\   \hline
 $D\pi$                  &  $9 \pm 8$        & $0 \pm 5$                  & $6 \pm 4$ \\   \hline
\end{tabular}
\end{center}
\caption{Estimates of the numbers of background events in the $m_{ES}>$ 5.272 GeV$/c^2$  region from the fit to data in the full $m_{ES}$  region  for the   $ B^- \to D^0 K^-$, $B^- \to D^{*0}(D^0 \pi^0) K^-$, and $B^- \to D^{*0}(D^0\gamma) K^-$ samples.}
\label{tab:bgfrac}
\end{table}

\begin{table}[h]
\begin{center}
\begin{tabular}{|c|c|c|c|}
\hline
 $q \bar q$ background parameters    &  $D^0 K^-$       & $(D^0 \pi^0) K^-$   & $(D^0\gamma) K^-$ \\   \hline\hline
 $f_{D^0}$               &  $0.27\pm 0.06$  & $0.27\pm 0.13$    & $0.13\pm 0.02$  \\   \hline
 $R$                     &  $0.21\pm 0.03$  & $0.23\pm 0.11$    & $0.16\pm 0.06$  \\   \hline
\end{tabular}
\end{center}
\caption{$q \bar q$ background parameters  for the  $ B^- \to D^0 K^-$, $B^- \to D^{*0}(D^0 \pi^0) K^-$, and $B^- \to D^{*0}(D^0\gamma) K^-$ samples.}
\label{tab:bgparams}
\end{table}

\subsection{Likelihood fit on  control samples} 

 We test our $CP$ fit procedure on two high statistics control samples:  $D^{*+}\rightarrow D^0\pi^+$ from $c \bar c $ continuum events and  $B^- \rightarrow D^{(*)0}\pi^-$. 
 The $D^{*+}\rightarrow D^0\pi^+$ sample mimics  a $B^- \rightarrow D^0 K^-$ sample with \rb $= 0$. The $B^- \rightarrow D^0\pi^-$ sample is similar to  $B^- \rightarrow D^0 K^-$, but its \rb~  is expected to be approximately 0.007 \cite{dspaper}.
In the $CP$ fit to  $D^{*+}\rightarrow D^0\pi^+$  we obtain $r_B=(-5.2\pm5.2)\times 10^{-3}$. In the $B^- \rightarrow D^{0}\pi^-$  we obtain   \rb$= (1.8\pm1.5)\times 10^{-2} $, $\gamma = (18\pm45)^\circ$, $\delta_B = (246\pm43)^\circ$
  and  in  $B^-\rightarrow D^{*0}\pi^-$ we find \rbs$= (4.6\pm2.1)\times 10^{-2}$, $\gamma  = (90\pm35)^\circ$, $\delta^*_B = (117\pm35)^\circ$.
  The results obtained are consistent with the expectations  of Monte Carlo  experiments.

%% file: results-prl.tex
\subsection{ Likelihood fit on $B^- \to D^{(*)0} K^-$ sample}

\begin{figure}[btph]
\begin{center}
\includegraphics[height=7cm]{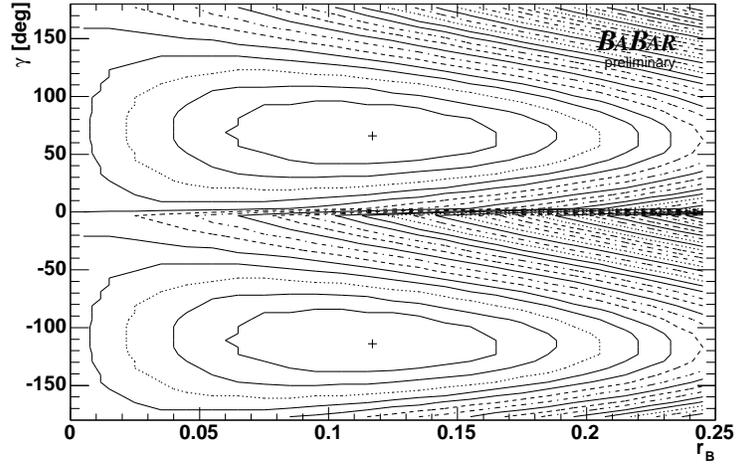}
\includegraphics[height=7cm]{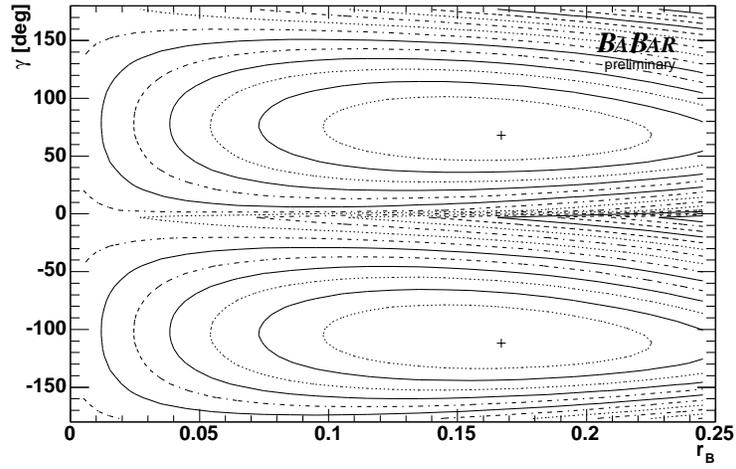}
\caption{ $\ln{\cal L}$ contour plots in $\gamma$ versus \rb~  in  $B^- \ra D^0 K^-$  (top) and   $B^- \ra D^{*0} K^-$ (bottom).  $\delta_B$ is fixed to the value obtained from the fit. Each contour represent a $\ln{\cal L}$ variation of 0.5.      }
\label{fig:likecontour}
\end{center}
\end{figure}

\begin{figure}[btph]
\begin{center}
\includegraphics[height=7cm]{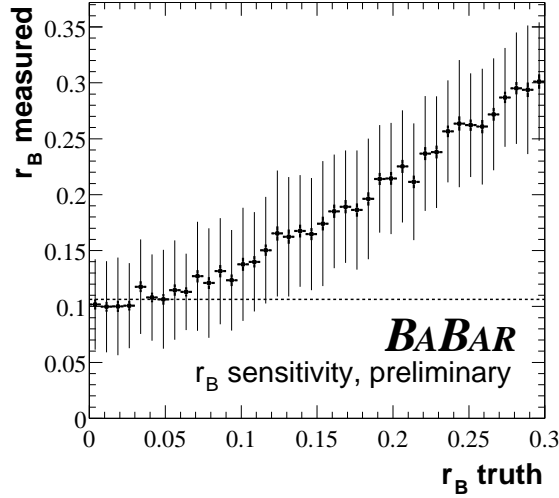}
\caption{ \rb~ values obtained in the likelihood fit versus generated \rb~ values in Monte Carlo $B^- \ra D^0 K^-$ experiments. The error bars represent  the RMS of the \rb~ values returned by the fit. The dashed horizontal line indicates the \rb~ value ($0.106$)  found in data.    }
\label{fig:toymcbias}
\end{center}
\end{figure}

In the sample of 227 million $B\bar B$ events   we obtain the following  signal yields 
\begin{eqnarray}
   N(B^- \ra D^0 K^-)                 & = & 282 \pm 20 ,\\ \nonumber
   N(B^- \ra D^{*0}(D^0 \pi^0) K^-)   & = & 89 \pm  11 ,\\ \nonumber
   N(B^- \ra D^{*0}(D^0 \gamma) K^-)  & = & 44  \pm 8 ,
\end{eqnarray}
 in agreement with our expectation from simulation and measured branching ratios.
 We obtain the following $CP$ parameters from the fit for $B^- \ra D^0 K^-$,  \rb$ =0.117 \pm 0.053$, $\delta_B=(109 \pm 28)^\circ$, and $ \gamma=(66 \pm 28)^\circ$  and for  $B^- \ra D^{*0} K^-$, \rbs$ =0.167 \pm 0.065$, $\delta^*_B=(294 \pm 28)^\circ$, and $\gamma=(68 \pm 29)^\circ$. These  errors are estimated with a Gaussian assumption for the likelihood. 
 However,  for this small sample, these  low \rb~ and \rbs~ fitted values lead   to a non-Gaussian behavior  of the likelihood function as shown in  Fig.~\ref{fig:likecontour},  necessitating  a different  approach, described next, in the computation of the   confidence intervals for \rb , $\gamma$ and $\delta_B$. Fig.~\ref{fig:toymcbias} shows for \rb~ values generated in the $[0.0,0.3]$ range the \rb~ values obtained in fits to Monte Carlo experiments of the same size as data. While the $CP$ fit is linear for large values of \rb, it is not sensitive to \rb~ values below 0.1.   This problem did not exist for the larger   $D^{*+}\rightarrow D^0\pi^+$  and  $B^- \rightarrow D^{(*)0} \pi^-$ samples as  we have verified using Monte Carlo simulation. 

\begin{figure}[btph]
\begin{center}
\includegraphics[height=5cm]{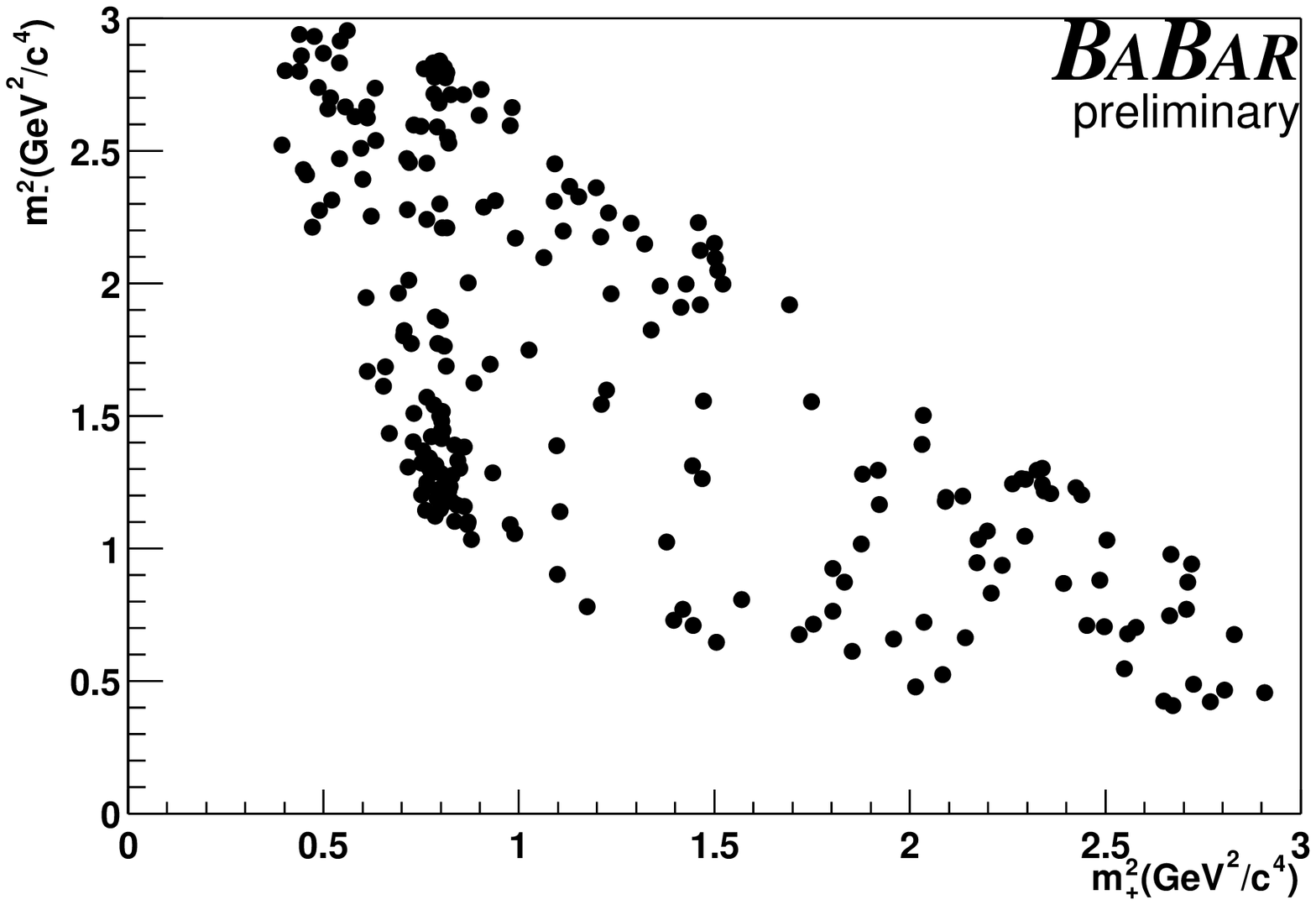}
\includegraphics[height=5cm]{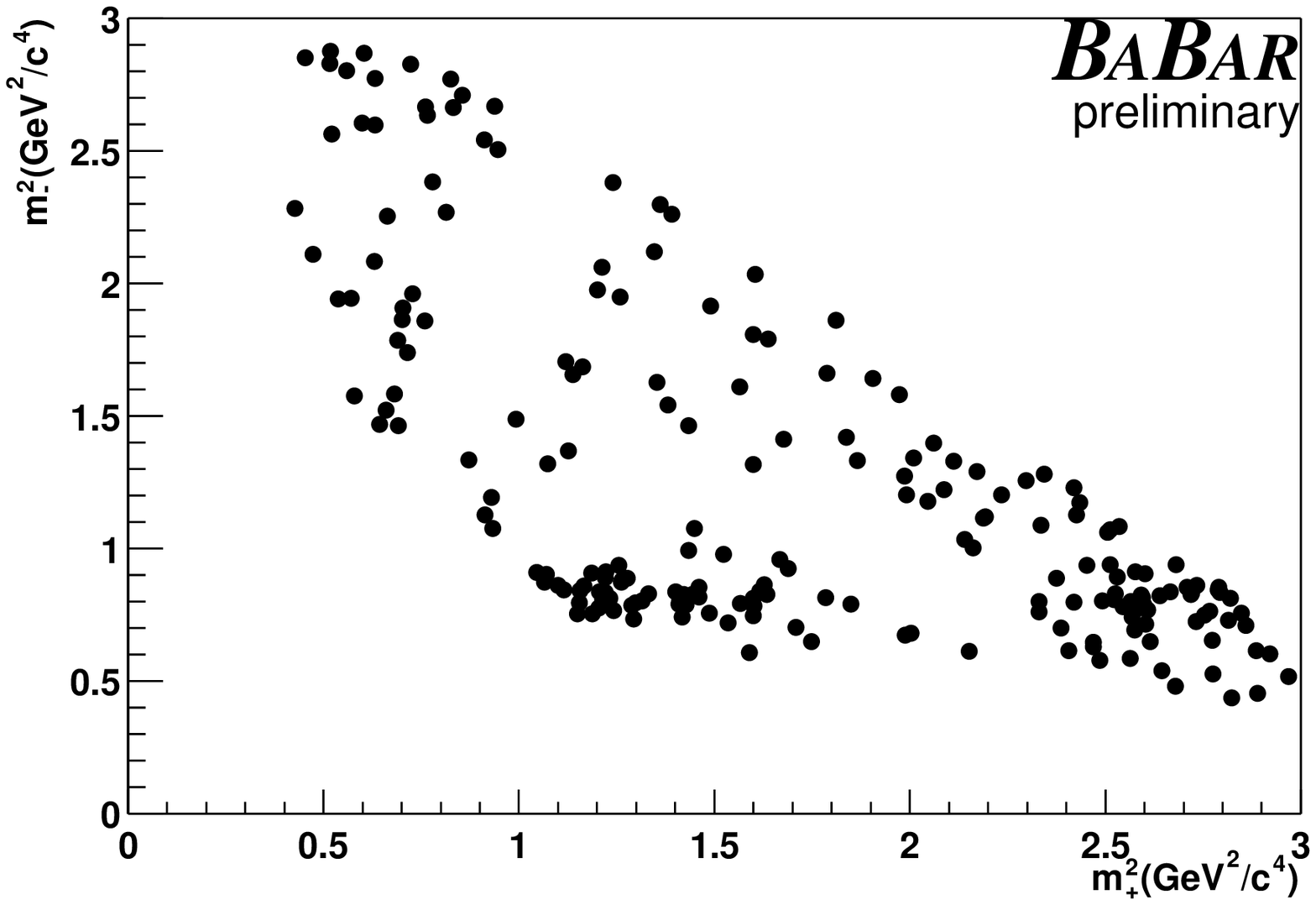}
\includegraphics[height=5cm]{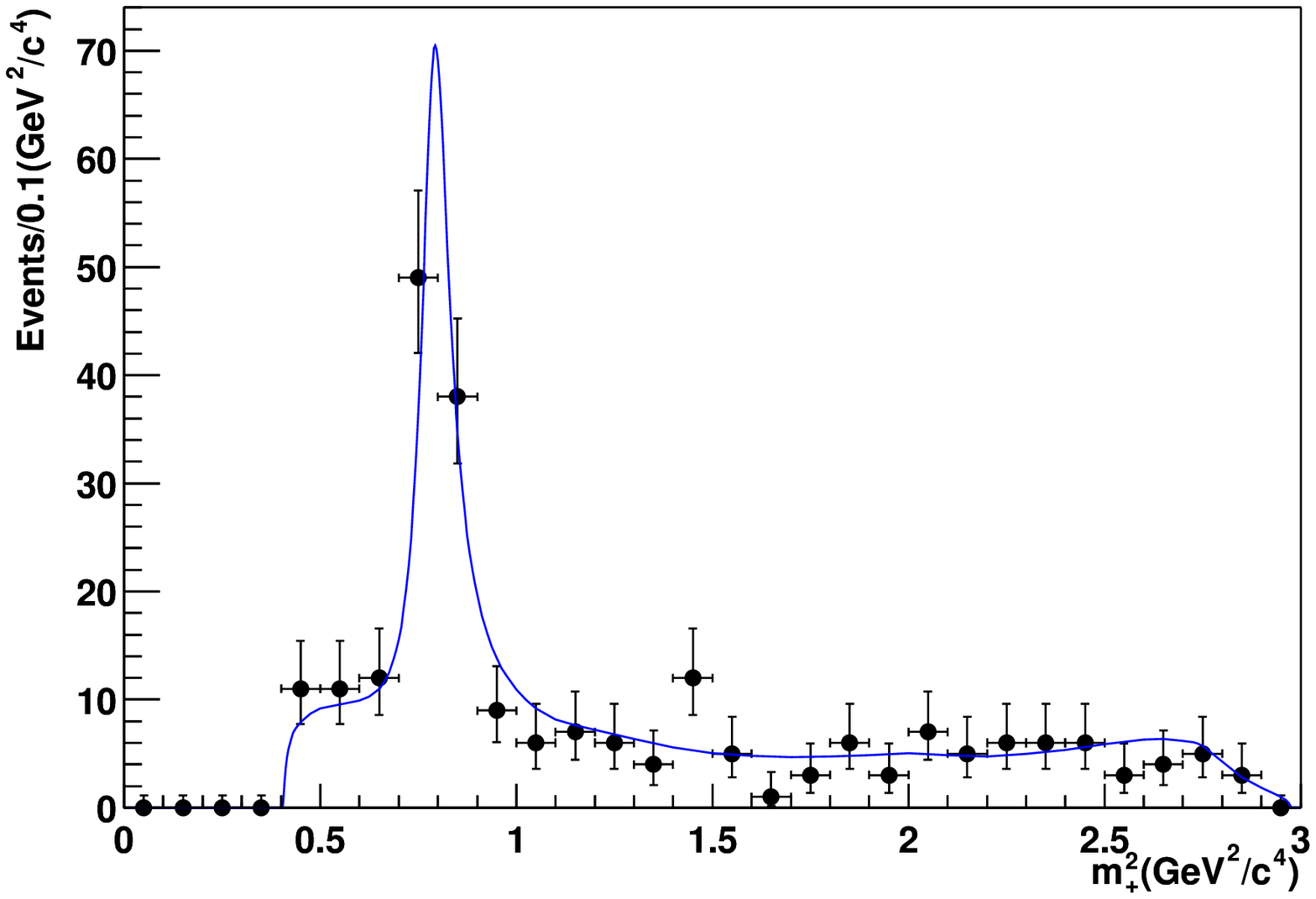}
\includegraphics[height=5cm]{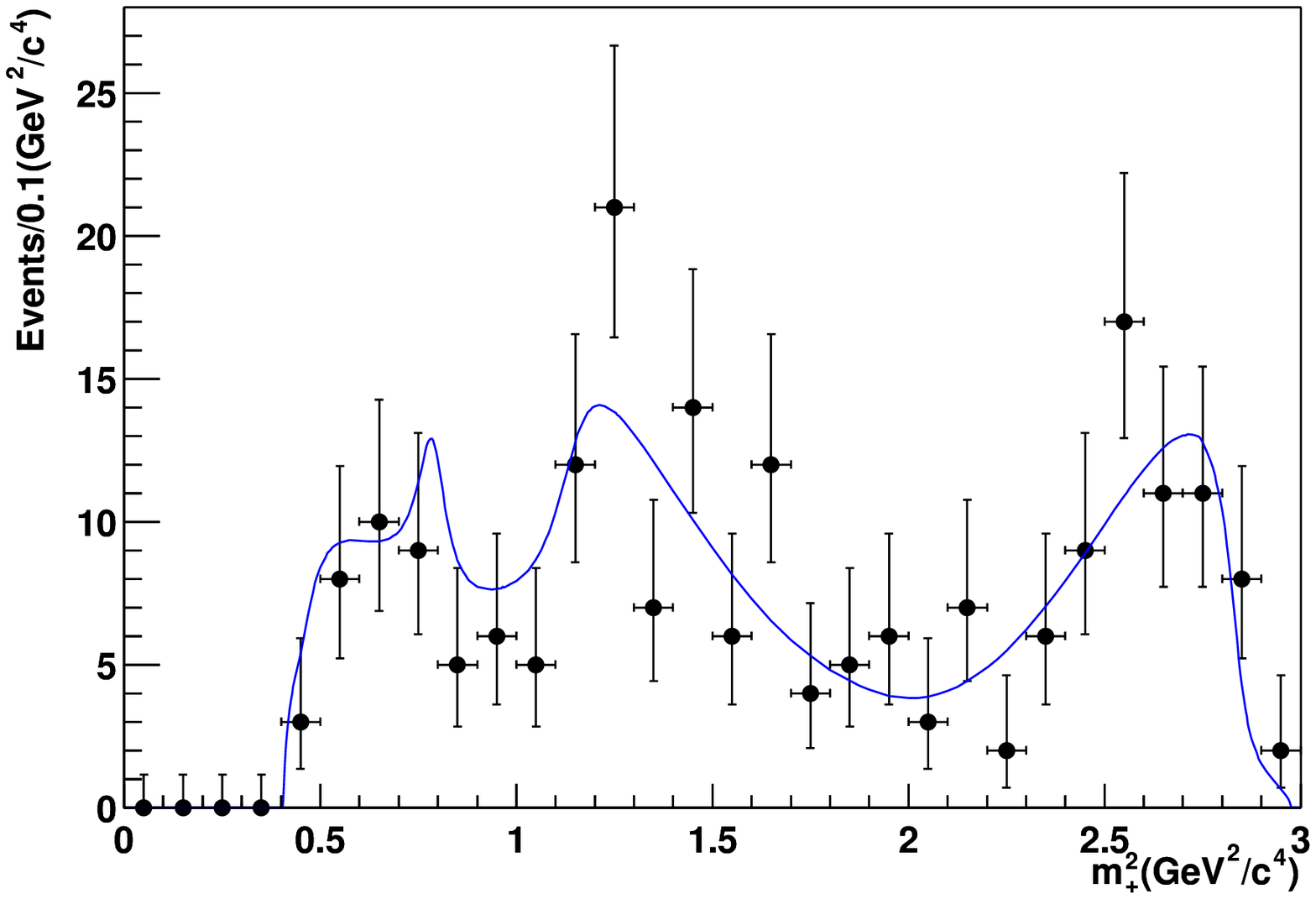}
\includegraphics[height=5cm]{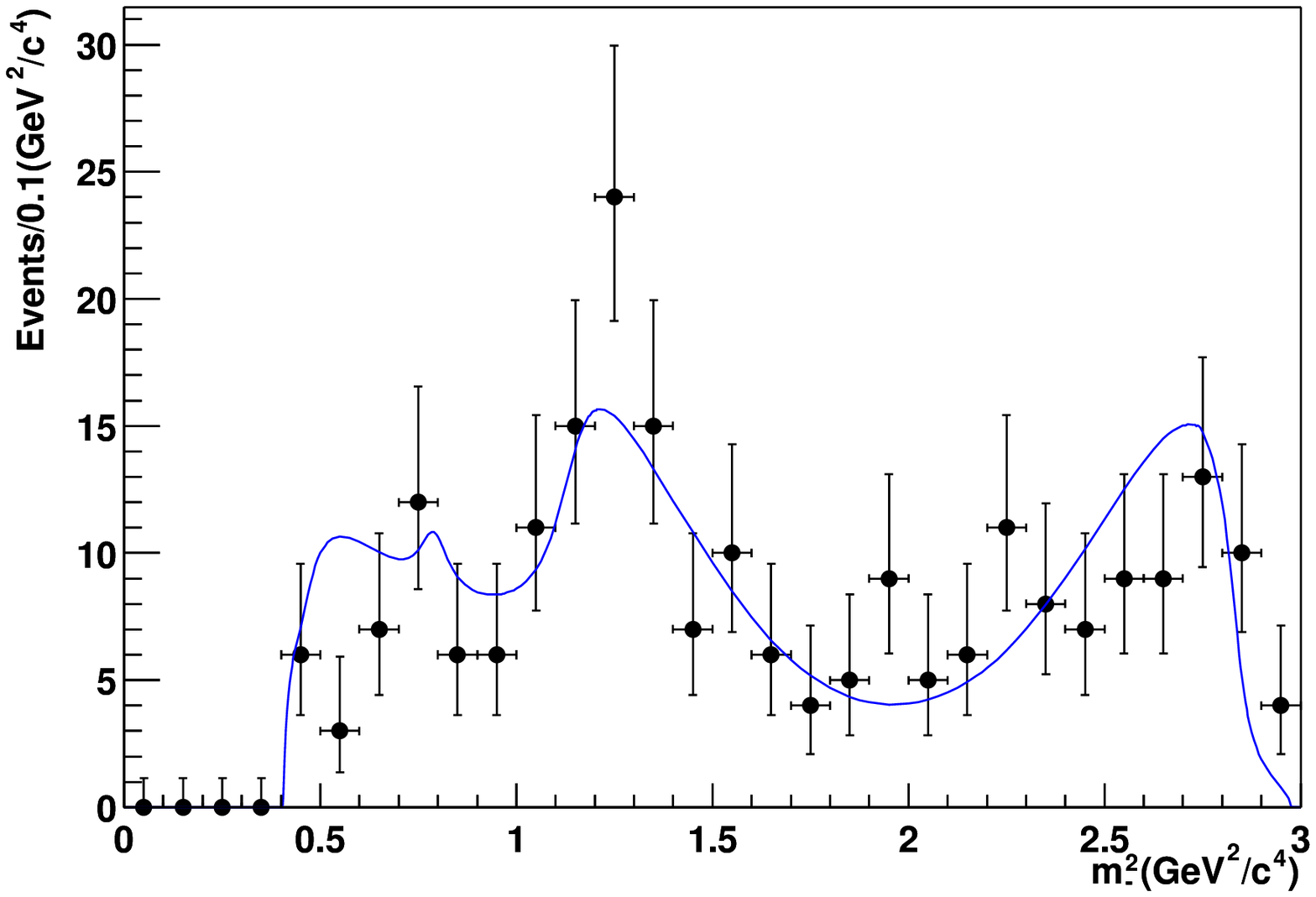}
\includegraphics[height=5cm]{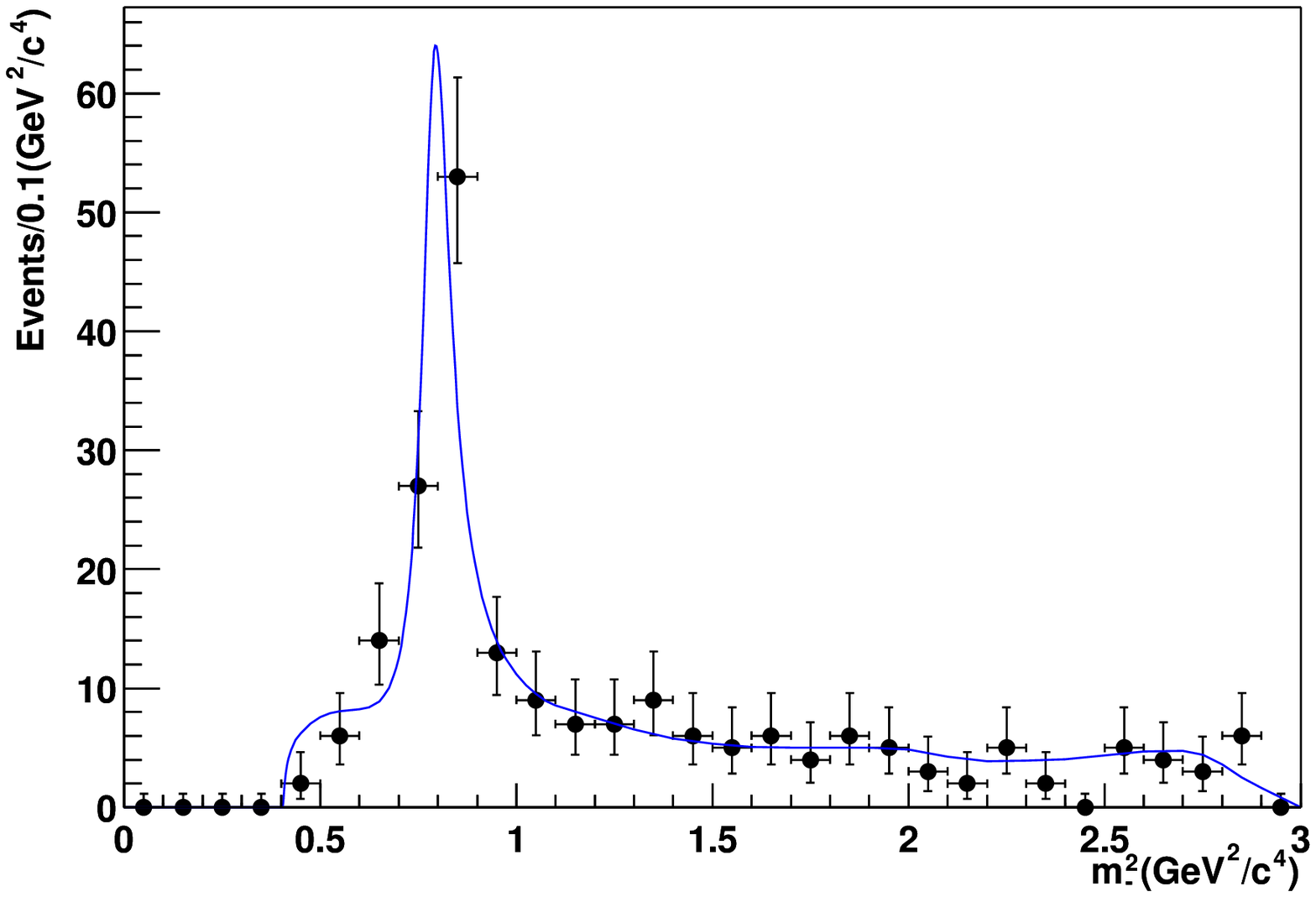}
\caption{ Dalitz distribution (first row) from the  $B^- \ra D^0 K^-$ events with  $m_{ES}  > 5.272$ GeV$/c^2$. Projections on $m^2_+$ (second row) and $m^2_-$ (third row) are shown with the result of the fit  superimposed. In the left column   $B^+$ candidates are shown, in the right  $B^-$ candidates.}
\label{fig:DKprojdalitz}
\end{center}
\end{figure}

\begin{figure}[btph]
\begin{center}
\includegraphics[height=5cm]{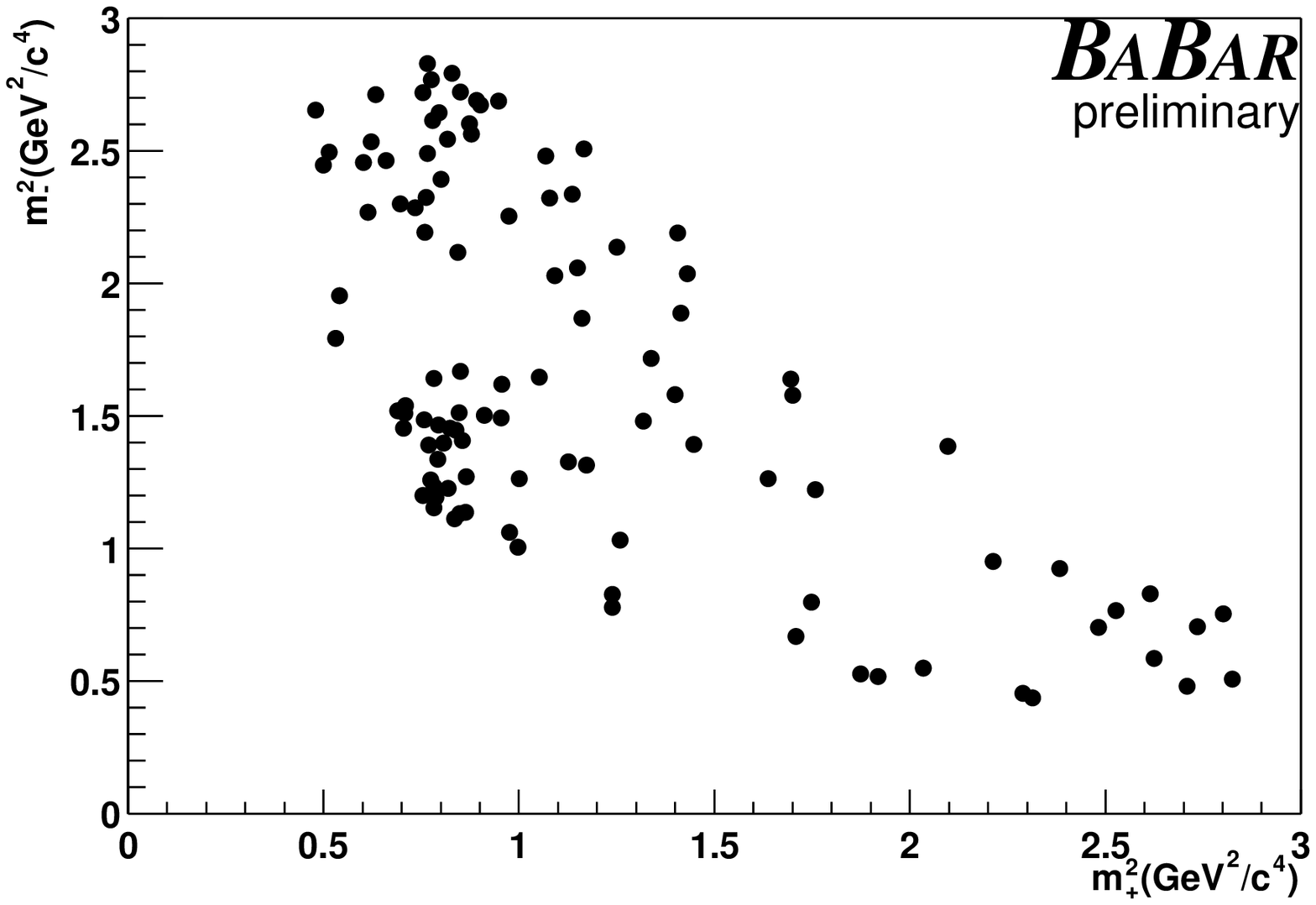}
\includegraphics[height=5cm]{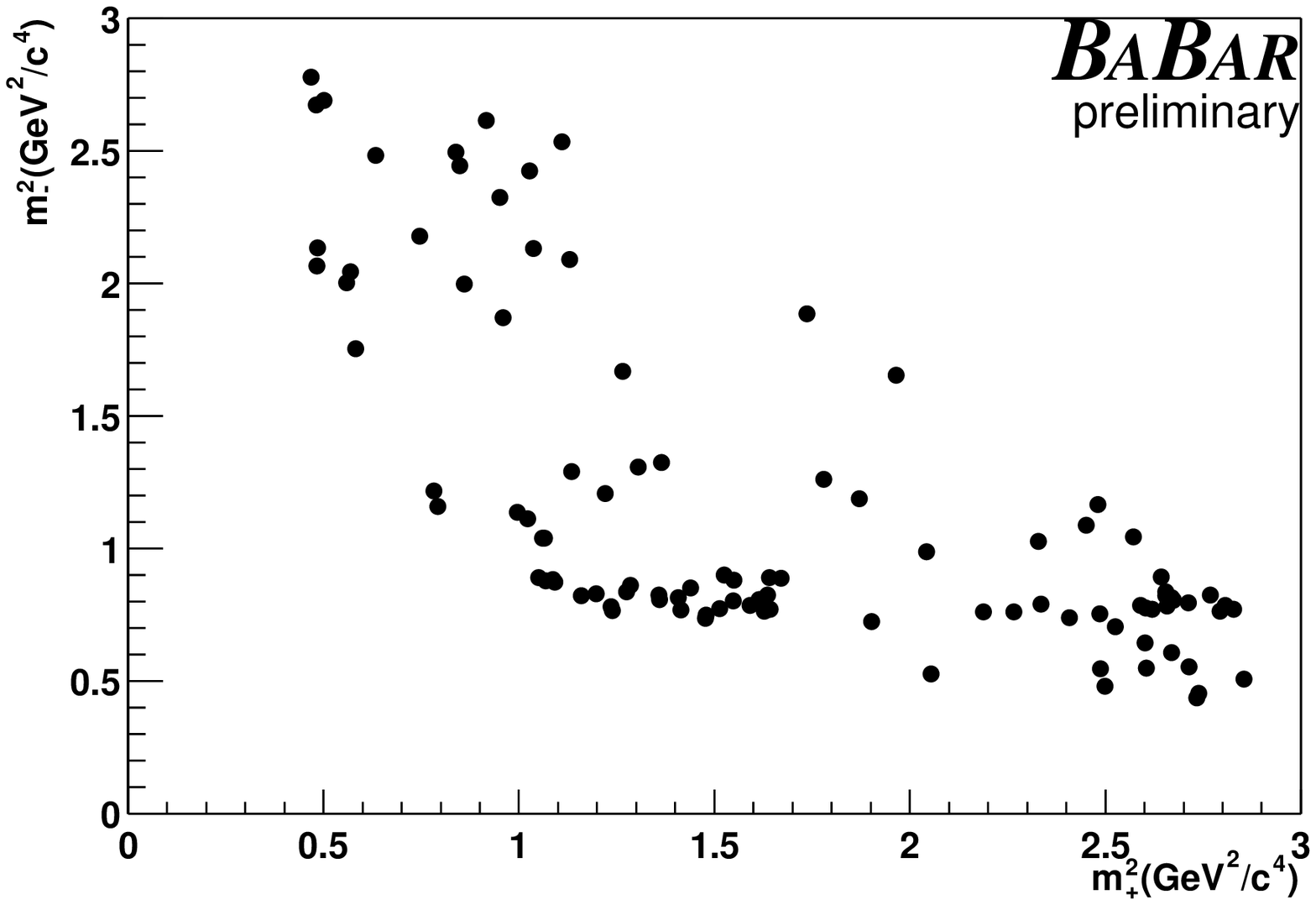}
\includegraphics[height=5cm]{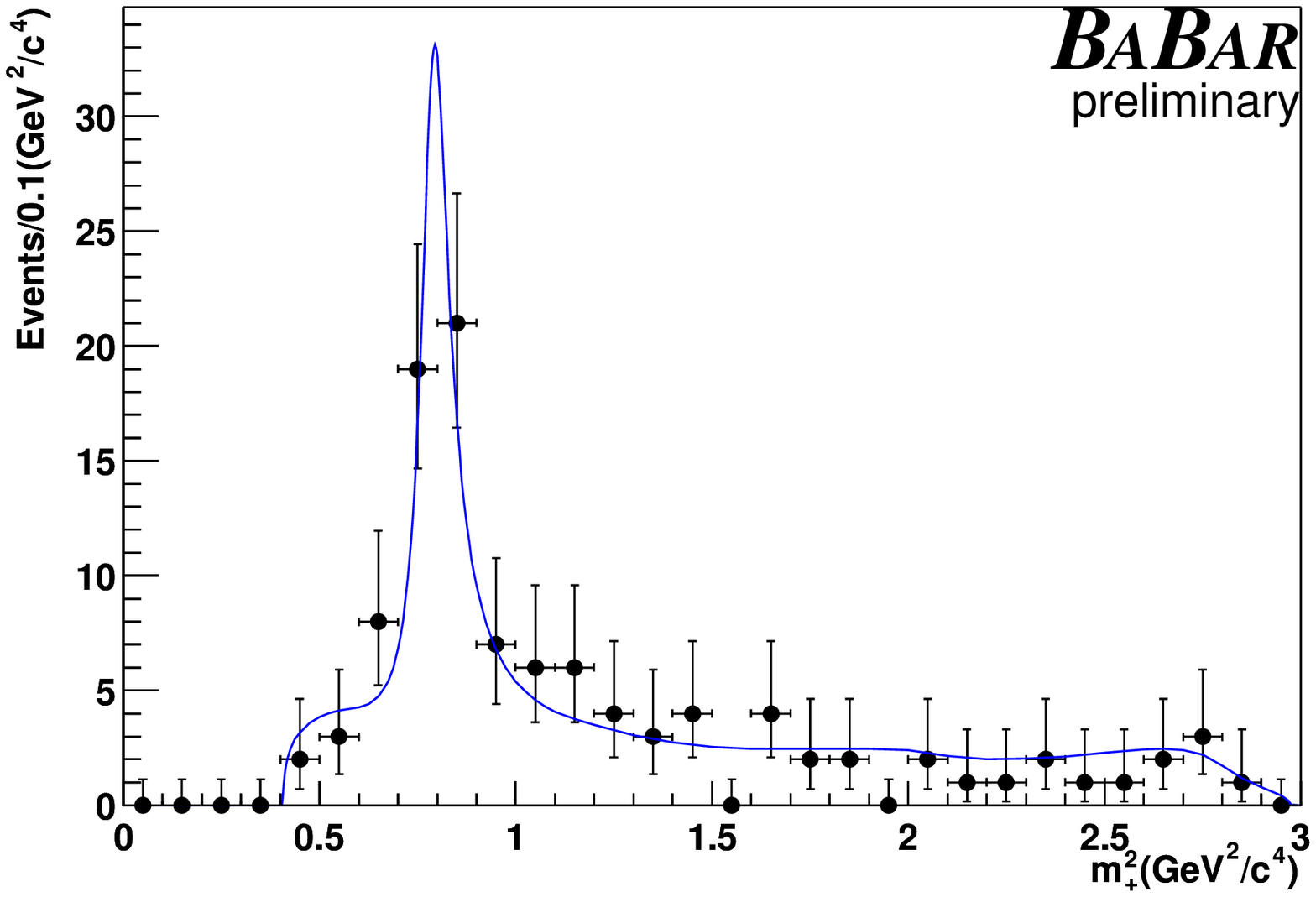}
\includegraphics[height=5cm]{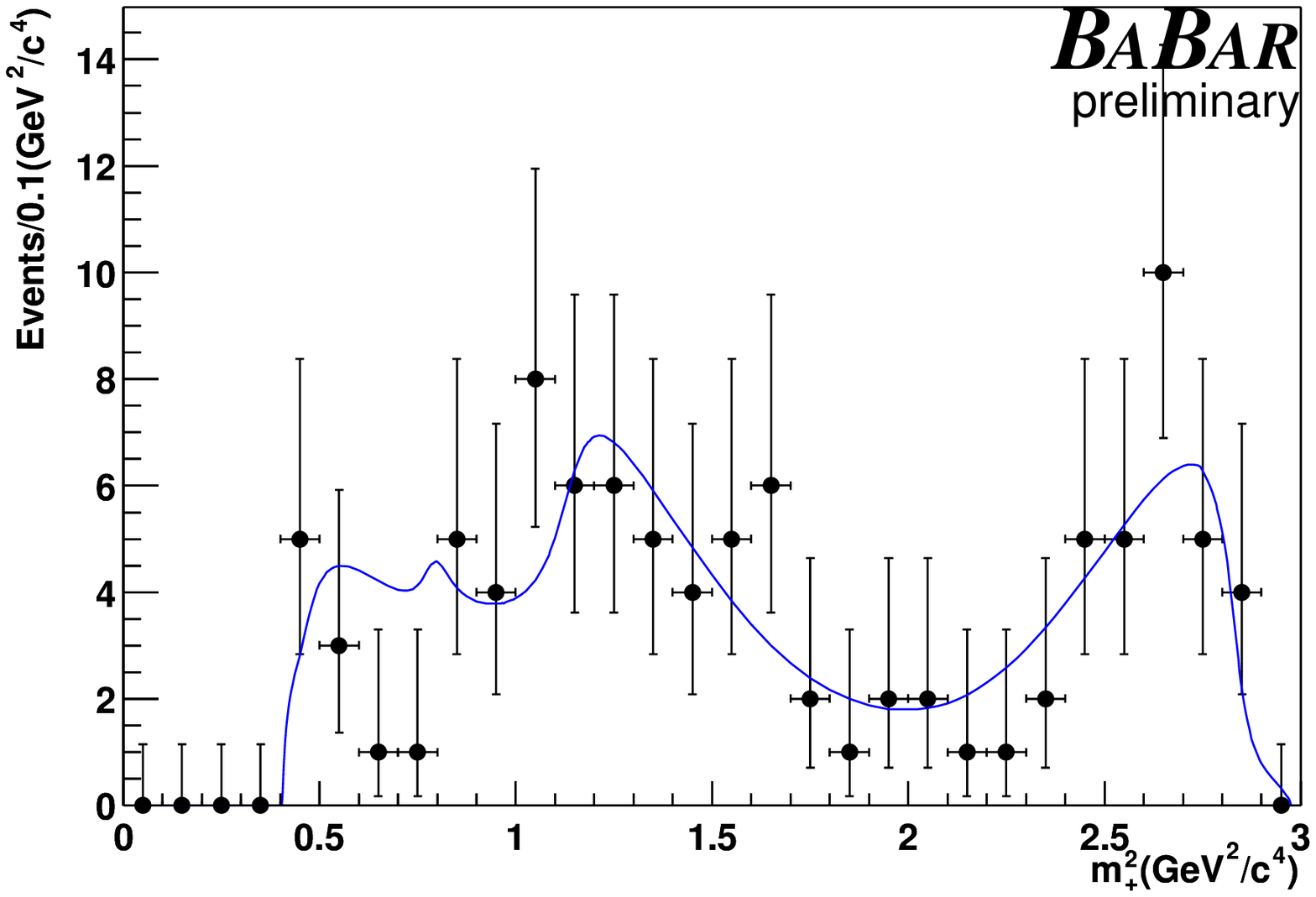}
\includegraphics[height=5cm]{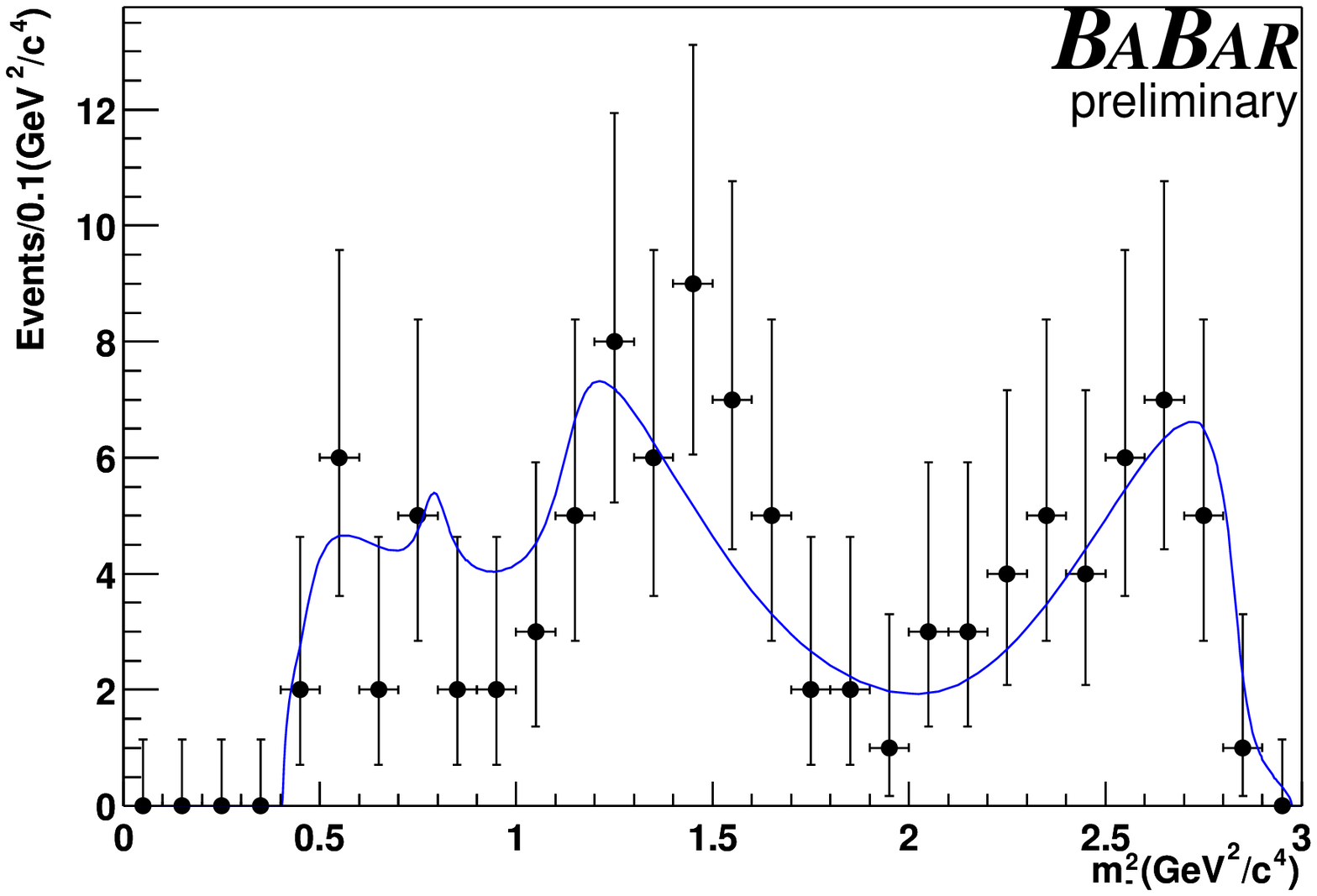}
\includegraphics[height=5cm]{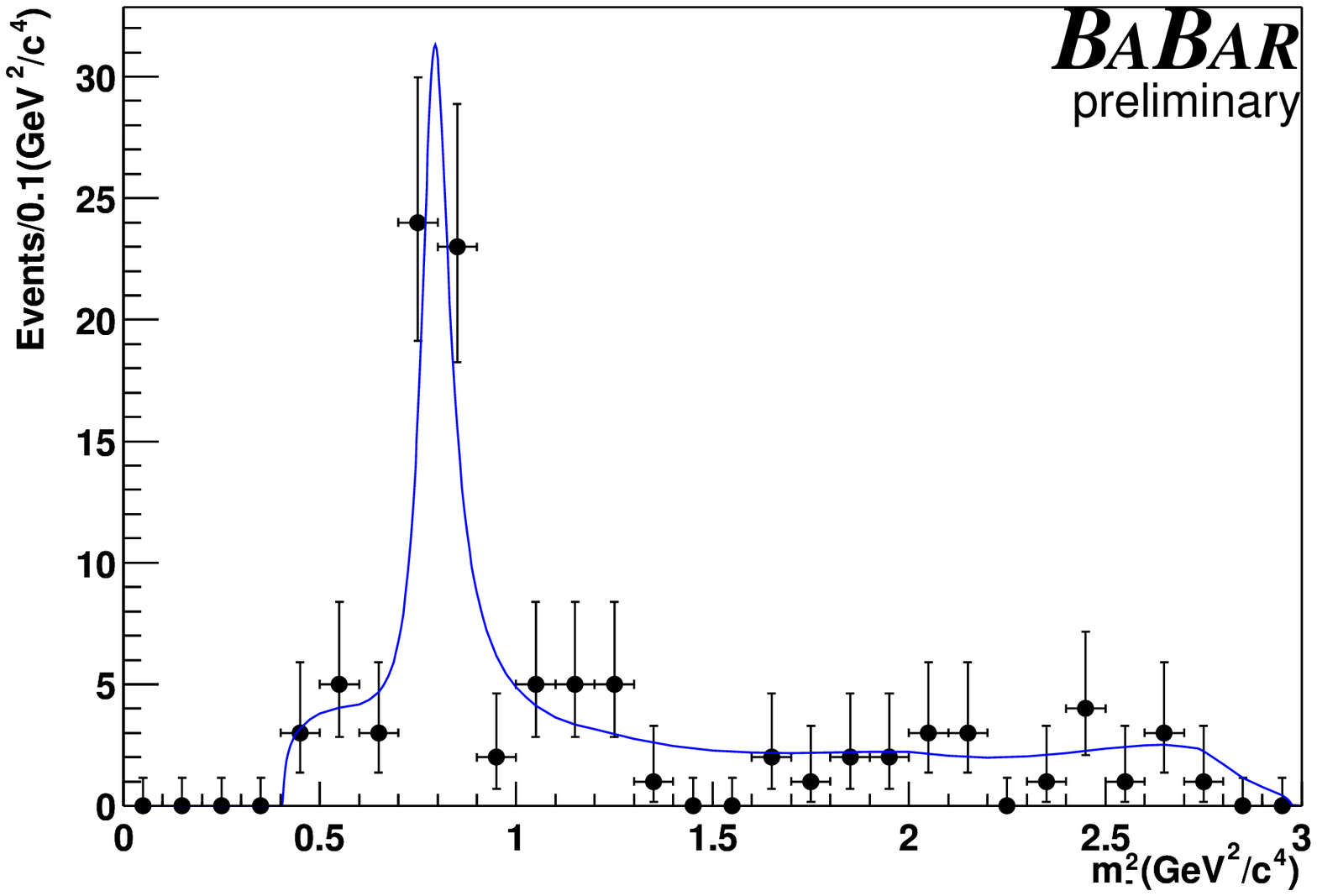}
\caption{  Dalitz distribution (first row) from the  $B^- \ra D^{*0} K^-$ events with  $m_{ES}  > 5.272$ GeV$/c^2$. Projections on $m^2_+$ (second row) and $m^2_-$ (third row) are shown with the result of the fit  superimposed. In the left column   $B^+$ candidates are shown, in the right  $B^-$ candidates.  }
\label{fig:DstKprojdalitz}
\end{center}
\end{figure}

 In Fig.~\ref{fig:DKprojdalitz} and  Fig.~\ref{fig:DstKprojdalitz} we show the Dalitz distribution and the  $m^2_+$ and $m^2_-$ projections for events with $m_{ES} > 5.272$ GeV$/c^2$ for $B^- \ra D^0 K^-$ and  $B^- \ra D^{*0} K^-$ respectively.   $B^+$ and $B^-$ candidates  distributions are separately shown  with the total PDF superimposed.

\subsection{Confidence intervals of the $CP$ parameters}

 We evaluate  the  likelihood function  $\cal{L}$ (\rb, $\gamma$, $\delta_B$) after fixing all  parameters except (\rb, $\gamma$, $\delta_B$) which are varied   in their  range of definition,   [0,1], [-$\pi$,$\pi$], and [0,2$\pi$], respectively. We then estimate the confidence region for the $CP$ parameters using a Bayesian technique. This implies a choice of a priori distribution. For this preliminary result we arbitrarily assume a uniform a priori distribution for each of the $CP$ parameters \rb,$\gamma$ and $\delta_B$. 
 
  In $\gamma$-\rb~ space  we define a two-dimensional confidence region $\cal{D}(\cal{C})$ corresponding to  a given confidence level  $\cal{C} $  

 \begin{center}
 \begin{math}
  \frac{ \int_{D(\cal{C})} d r_B d \gamma  \int_0^{2\pi}d \delta_B  {\cal L} ( r_B,\gamma,\delta_B) }{\int_0^{1}d r_B \int_{-\pi}^
{\pi}d\gamma \int_0^{2\pi}d\delta_B {\cal L}( r_B ,\gamma,\delta_B)\,  } =  \cal{C}
 \end{math}
\end{center}
 We uniquely define   $D(\cal{C})$  by requiring that the likelihood value at any point on the boundary of $\cal{D}$ be  the same and integrating over all likelihood values larger than the value at the boundary.

 Similarly we  define  one dimensional confidence intervals $I(\cal{C})$ corresponding to a   confidence level  $\cal{C}$.   For example, the interval for \rb~ is defined as  
 \begin{center}
 \begin{math}
  \frac{ \int_{I(\cal{C})} d r_B  \int_{-\pi}^{+\pi}d \gamma  \int_0^{2\pi} d \delta_B  {\cal L}( r_B,\gamma,\delta_B)  }{\
\int_0^{1}d r_B \int_{-\pi}^{\pi}d\gamma \int_0^{2\pi}d\delta_B  {\cal L}( r_B ,\gamma,\delta_B)\,d r_B } = \cal{C}
 \end{math}
\end{center}
  
\begin{figure}[btph]
\begin{center}
\includegraphics[height=5cm]{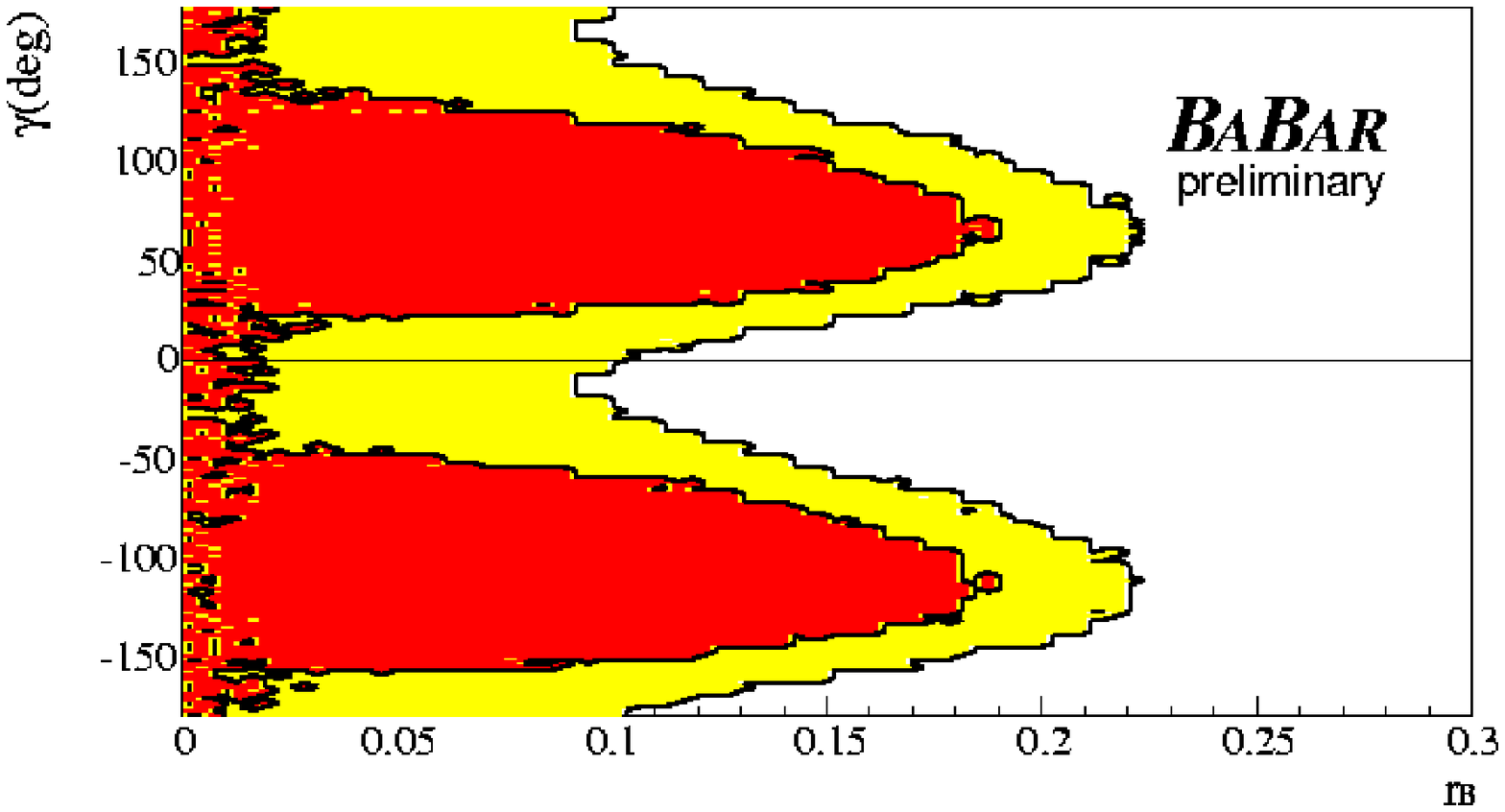}
\includegraphics[height=5cm]{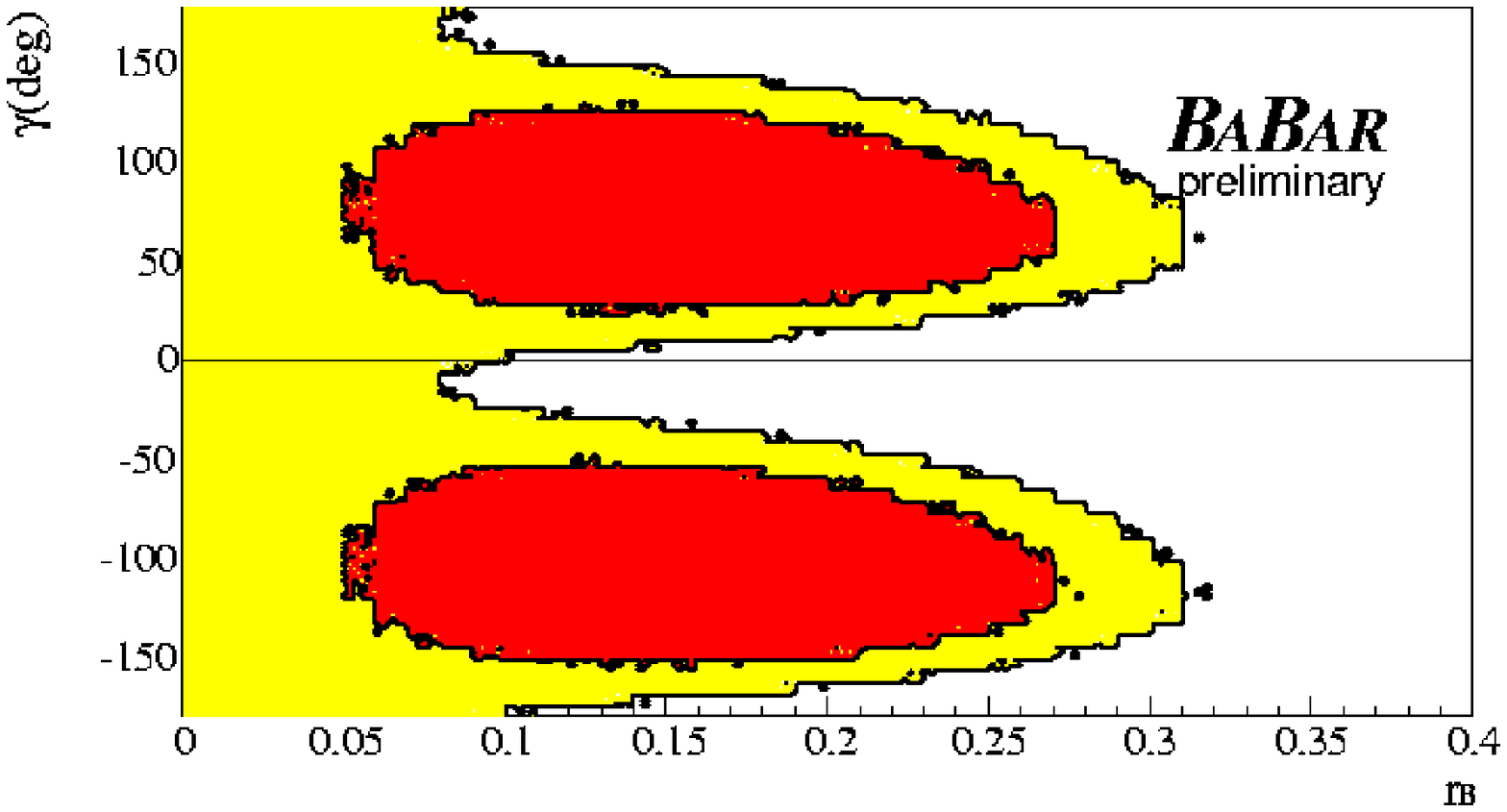}
\caption{ Bayesian confidence regions   for $\gamma$ versus \rb~ (top)  and \rbs~ (bottom)  for the $B^- \ra D^0K^-$ and $B^- \ra D^{*0}K^-$ samples respectively. The red (dark)  region corresponds to the 68\% confidence level region, the yellow (light) region to the 95\% C.L. region   }
\label{fig:DKDstargammarb}
\end{center}
\end{figure}

\begin{figure}[btph]
\begin{center}
\includegraphics[height=7cm]{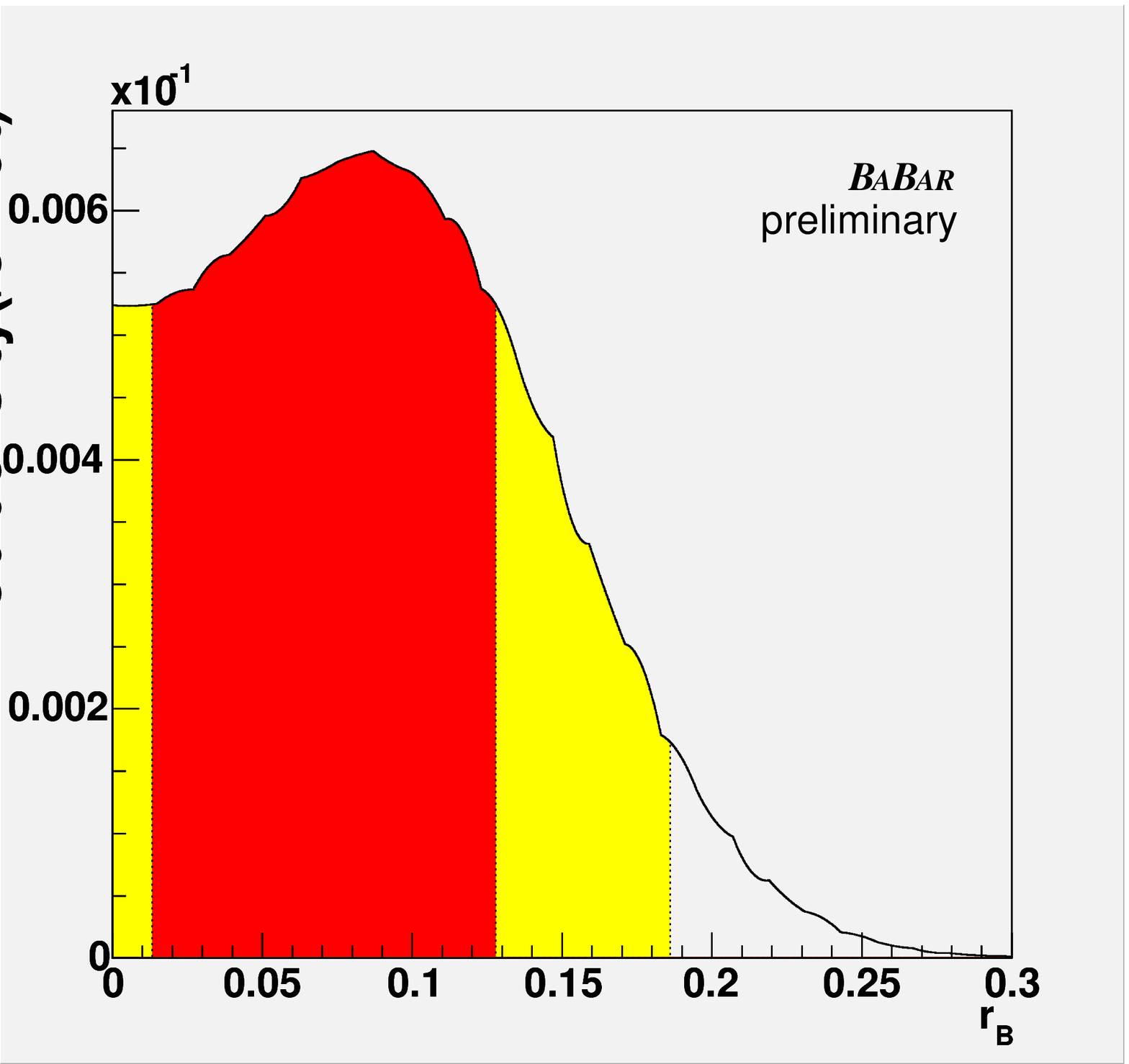}
\includegraphics[height=7cm]{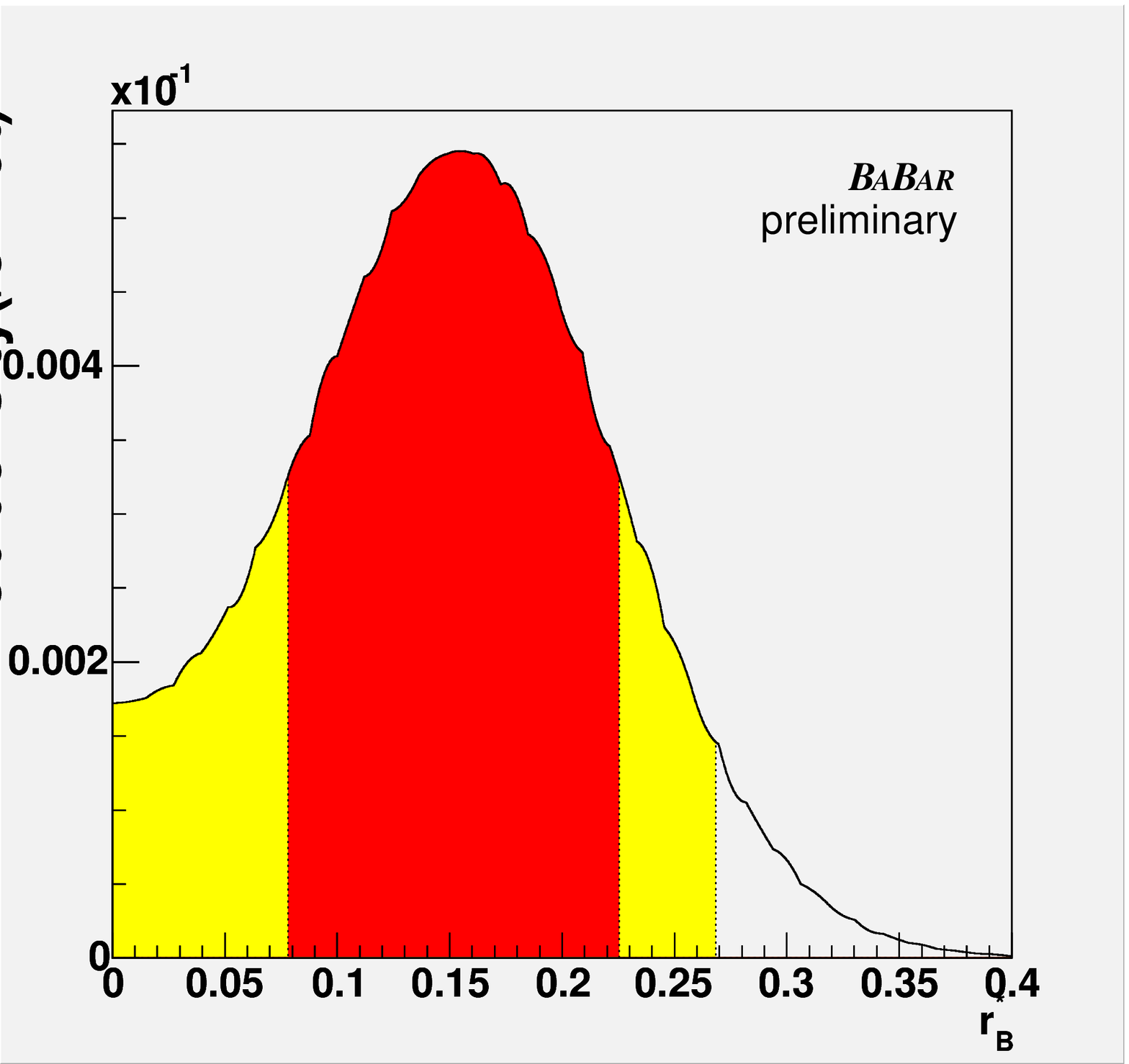}
\caption{ Probability density function for  \rb~ (left) and \rbs~(right). The red (dark)  region corresponds to the Bayesian 68\% confidence level  region, the yellow (light) region to the 95\% C.L. region.}
\label{fig:DKDstarrbpdf}
\end{center}
\end{figure}

\begin{figure}[btph]
\begin{center}
\includegraphics[height=7cm]{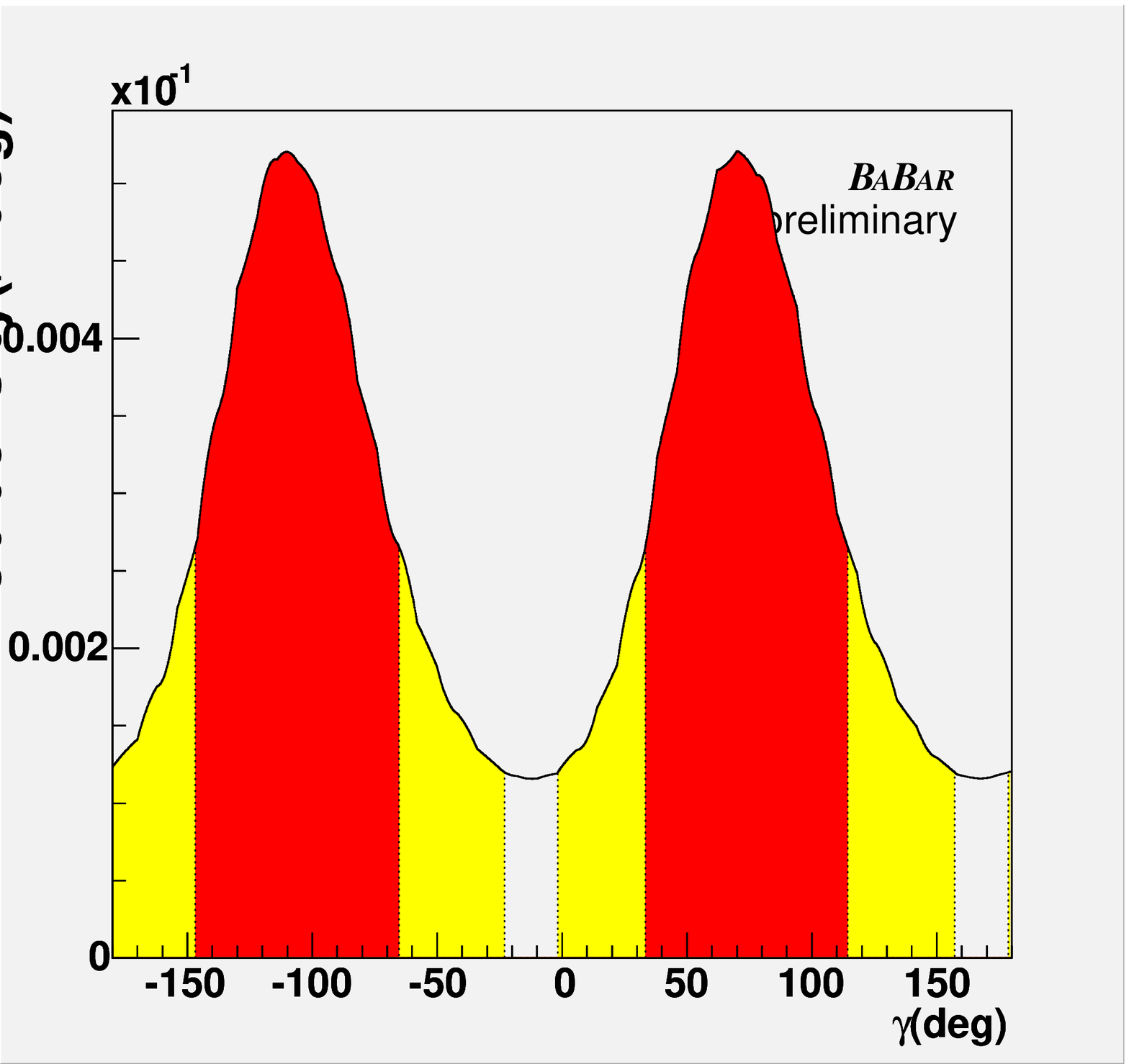}
\includegraphics[height=7cm]{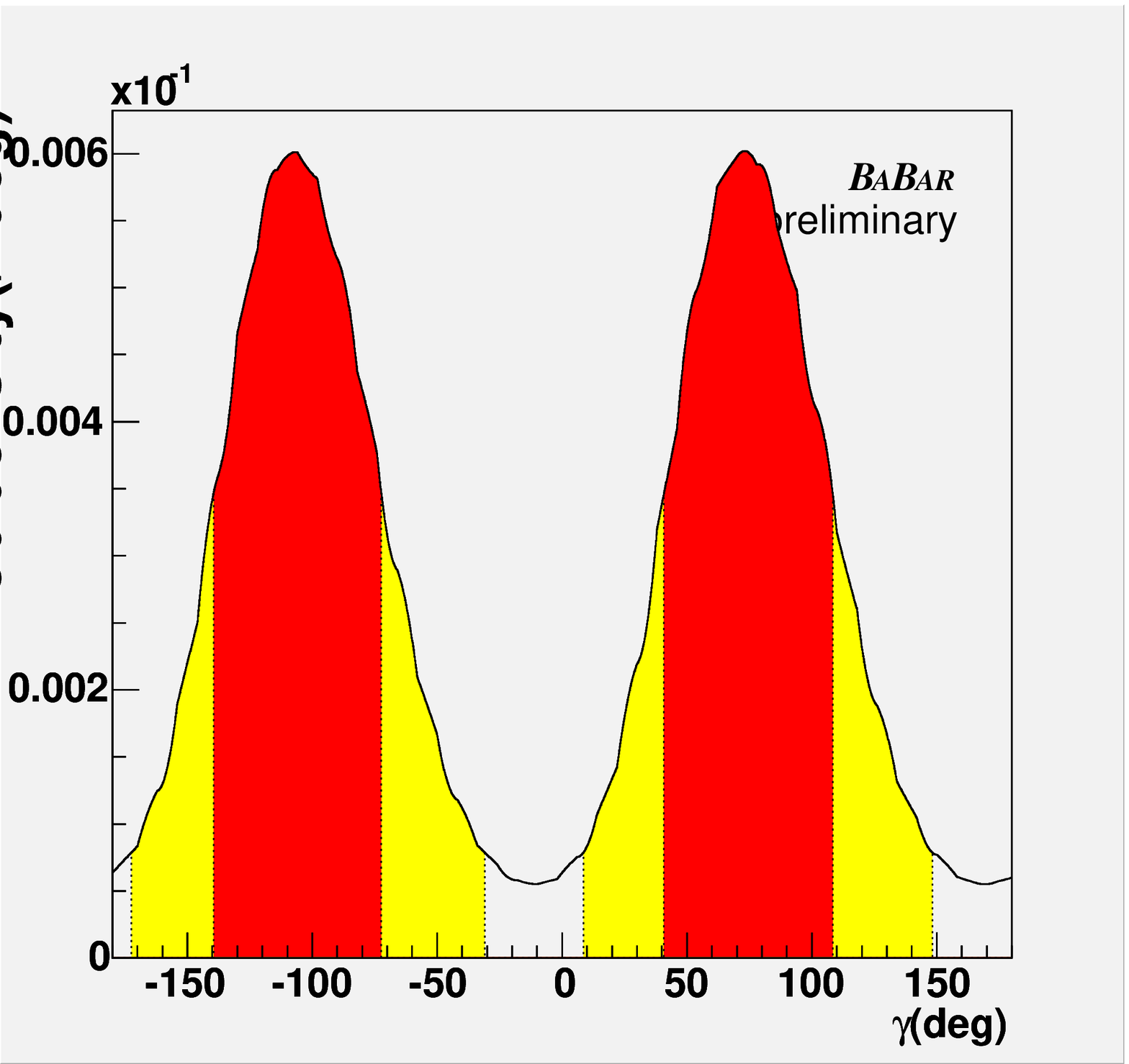}
\caption{ Probability density function for  $\gamma$  in $B^- \ra D^0K^-$ (left)  and $B^- \ra D^{*0}K^-$  (right) sample  . The red (dark)  region corresponds to the Bayesian 68\%   confidence interval region, the yellow (light) region to the 95\% C.L. region. The intervals for  $\gamma$ are disjoint  as a consequence of the  $\gamma \ra \gamma \pm \pi$ and  $\delta_B \ra \delta_B \pm \pi$ ambiguities.  }
\label{fig:DKDstargammapdf}
\end{center}
\end{figure}

\begin{figure}[btph]
\begin{center}
\includegraphics[height=7cm]{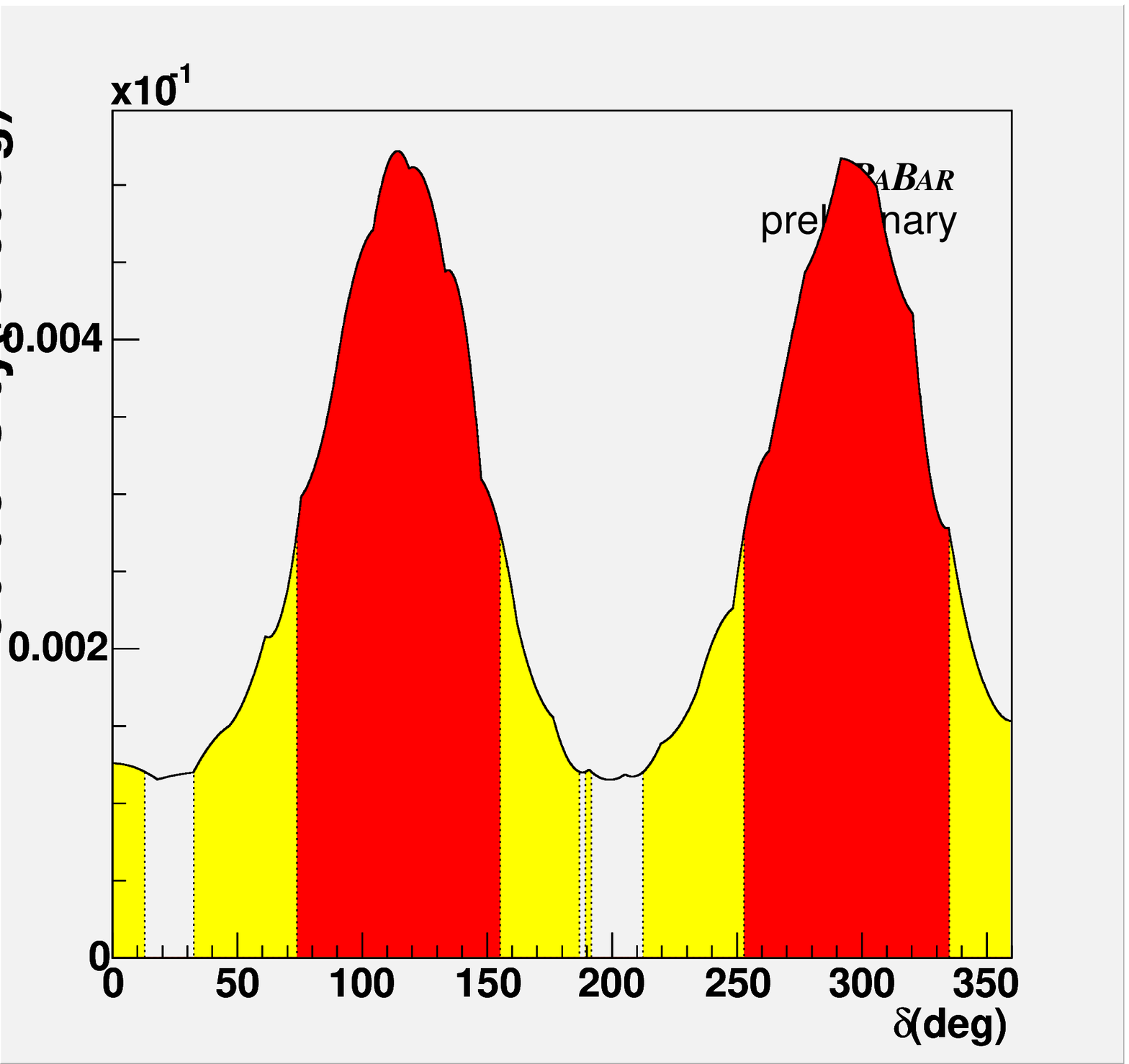}
\includegraphics[height=7cm]{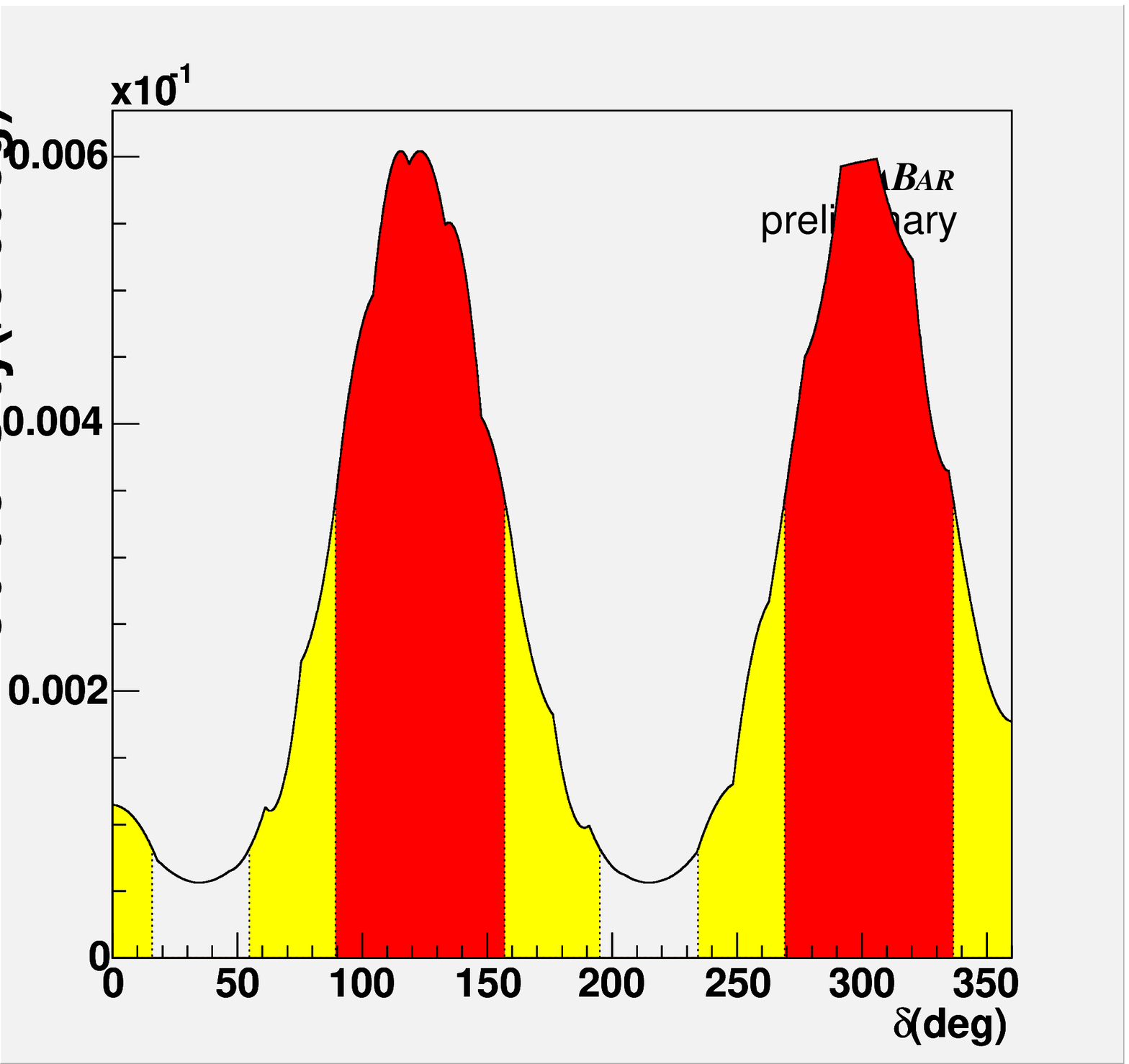}
\caption{ Probability density function for  $\delta_B$  in $B^- \ra D^0K^-$ (left) and $B^- \ra D^{*0}K^-$ (right)  sample. The red (dark)  region corresponds to the Bayesian 68\% confidence interval region, the yellow (light) region to the 95\% C.L. region. The intervals for $\delta_B$ are disjoint  as a consequence of the  $\gamma \ra \gamma \pm \pi$ and  $\delta_B \ra \delta_B \pm \pi$ ambiguities.  }
\label{fig:DKDstardeltapdf}
\end{center}
\end{figure}

 The two dimensional confidence regions $\cal{D}(\cal{C})$   in   $\gamma$ versus \rb~ and  $\gamma$ versus \rbs~  are shown in Fig.~\ref{fig:DKDstargammarb}. The red (dark) and yellow (light)  regions correspond  to the 68\% and 95\% confidence levels respectively.  The likelihood distributions for \rb, $\gamma$, and $\delta_B$  obtained by integrating  ${\cal L}$$($\rb$,\gamma,\delta_B)$ over the other two variables are shown in  Figs.~\ref{fig:DKDstarrbpdf}, \ref{fig:DKDstargammapdf} and \ref{fig:DKDstardeltapdf} respectively. The 68\% and 95\% confidence intervals are shown in red (dark) and yellow (light) respectively.
  The intervals for $\gamma$ and $\delta_B$ are disjoint  as a consequence of the $\gamma \ra \gamma \pm \pi$ and  $\delta_B \ra \delta_B \pm \pi$ ambiguities.
 
%% file: constraints.tex
\subsection{Constraints on \rb, $\delta_B$ and  $\gamma$ }

 From the procedure above we obtain Bayesian  68\% confidence intervals. We quote as  the central values   for $\gamma$ and  $\delta_B$  the average values weighted by  the likelihood distribution. The  errors associated with  the central values  are defined by  the boundaries of the  68\% confidence interval. We obtain  \rb$= 0.087 \pm ^{+0.041}_{-0.074},\,  \delta_B={114^\circ} \pm 41^\circ ({294^\circ} \pm 41^\circ ),\, \gamma= { 70^\circ}\pm 44^\circ  ({-110^\circ}\pm 44^\circ  )$ for $B^- \to D^0 K^-$ and 
 \rbs$ = 0.155 ^{+0.070}_{-0.077}  , \,  \delta^*_B ={303^\circ} \pm 34^\circ ({123^\circ} \pm 34^\circ),\,  \gamma={73^\circ} \pm 35^\circ ({-107^\circ} \pm 35^\circ )$ for $B^- \to D^{*0} K^-$.
\noindent
 We constrain \rb~ to be  $< 0.16$     at 90$\%$ confidence level.

\begin{figure}[btph]
\begin{center}
\includegraphics[height=7cm]{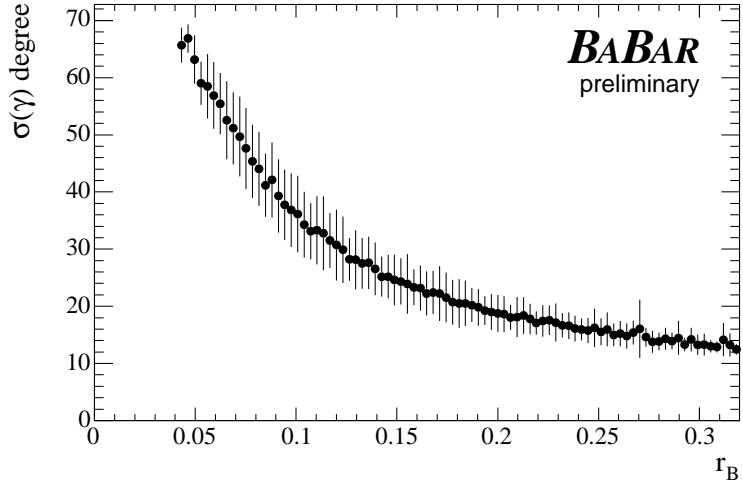}
\caption{ Statistical uncertainty in Gaussian hypothesis for $\gamma$ as a function of the value of \rb~  obtained  by the likelihood fit  in  Monte Carlo experiments.  }
\label{fig:sigmarbvsgamma}
\end{center}
\end{figure}

 As illustrated in Figs. \ref{fig:likecontour} and \ref{fig:sigmarbvsgamma}, the data have  no  sensitivity to $\gamma$ and $\delta_B$ for small values of \rb. Those values   are not excluded by our data.

\begin{figure}[btph]
\begin{center}
\includegraphics[height=7cm]{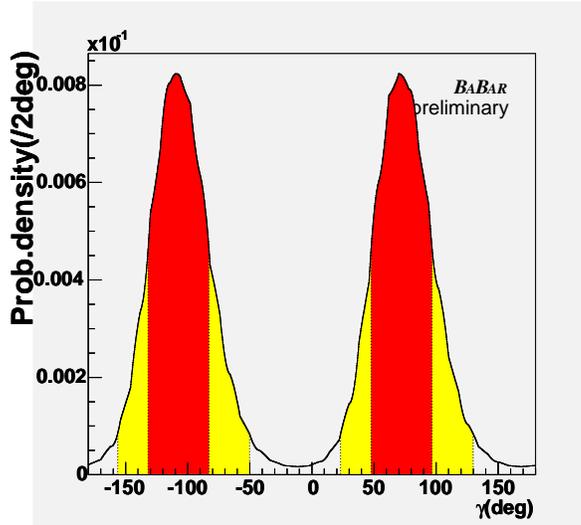}
\caption{ Probability density function for $\gamma$ from the combined samples.The red (dark)  region corresponds to the Bayesian 68\% confidence interval region, the yellow (light) region to the 95\% C.L. region.  }
\label{fig:gammacombined}
\end{center}
\end{figure}

We  construct  a combined likelihood function from the  product of the individual likelihoods  for $B^- \to D^0 K^-$  and $B^- \to D^{*0} K^-$  and we repeat   the  procedure outlined above.  From this we obtain $\gamma= {70^\circ} \pm 26^\circ$ (${-110^\circ} \pm 26^\circ$).  Fig.~\ref{fig:gammacombined}  shows the $\gamma$ likelihood distribution.

%% file: systematics-prl.tex
\begin{figure}[!htb]
\begin{center}
\includegraphics[height=12cm]{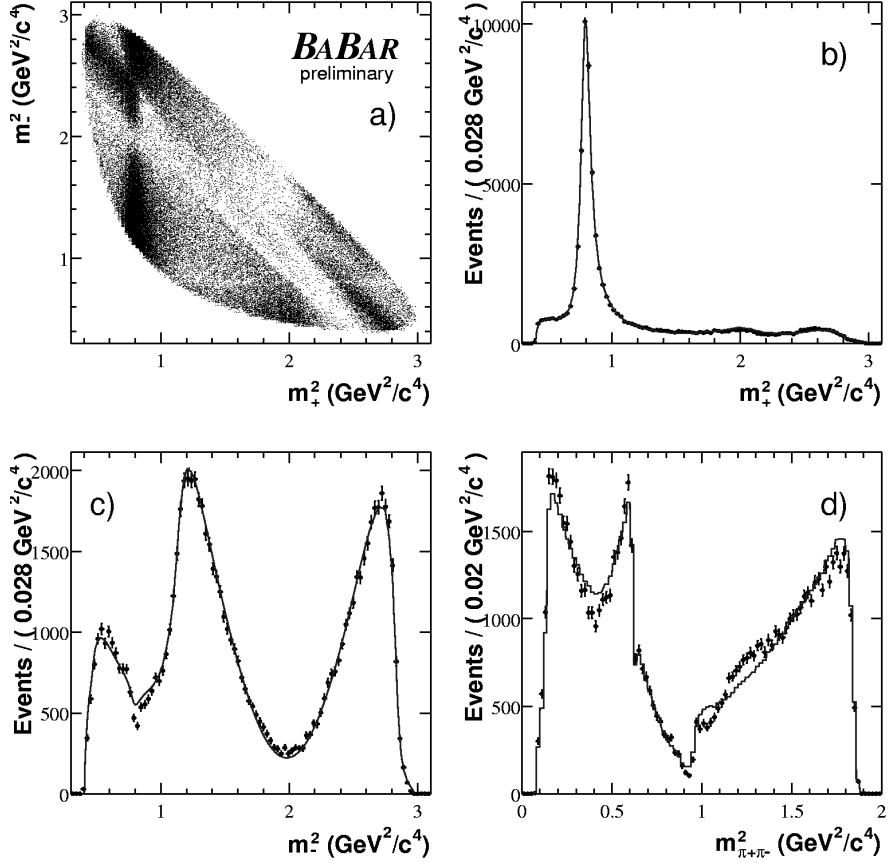}
\caption{CLEO model fit: a) the $\overline{D}^0 \ra K_S \pi^+ \pi^-$ Dalitz distribution from the $D^{*-} \ra \overline{D}^0 \pi^-$. Projection on (b) $m^2_+$,  (c) $m^2_-$  and (d) $m^2(\pi^+\pi^-)$   are  shown.   The result of the fit is superimposed and it is used to estimate the uncertainty due to the Dalitz model. }

\label{fig:cleomodel}
\end{center}
\end{figure}

The principal  systematic uncertainty on the measurement 
of $\gamma$ comes from the choice of the model used to describe 
 $D^0 \rightarrow K_S \pi^- \pi^+$ decay. We evaluate this uncertainty by considering  alternative models. For each model  we generate events and fit     both  the alternative model and the  nominal model (defined in Section \ref{sec:dalitzModel}) to these events. We quantify this uncertainty using  the differences in the fitted values for  $r_B$, $\delta_B$ and $\gamma$.
 For models where   the $\rho(1450)$, the $K^*(1680)$ 
and/or the doubly Cabibbo suppressed  $K^*_0$(1430) and $K^*_2$(1430) are removed 
or a different  description of resonances is  used,  the $\chi^2$ of the fit is not significantly different from  that of the  nominal model. 
 For these models the biases on $\gamma$ and $r_B$ are negligible and the
RMS of the distribution of the differences  is at most $1^{\circ}$ and 0.002 for $\gamma$ and $r_B$ respectively.
 As an extreme we  consider  a model  without the  $\sigma_1$ and/or $\sigma_2$ 
scalar,   or the CLEO Model \cite{ref:cleomodel}. Fits to these models result in  a significantly larger  $\chi^2$ than that of the  nominal model.  To illustrate the magnitude of this variation   Fig. \ref{fig:cleomodel}   shows   the result of a fit with the CLEO model. For these extreme  models the biases for $CP$ parameters are still small. The RMS of the differences for $\gamma$ and $r_B$ are 
approximately  $10^{\circ}$ and 0.02  respectively. 
 We conservatively assign   $\sigma_{\gamma}, \sigma_{\delta_B}  = 10^{\circ}$ and $\sigma_{\rb} = 0.02$  as systematic uncertainties associated with the Dalitz model.

 The summary of the estimates of other systematic uncertainties is given in Table \ref{tab:syst}.
 The most important effect is due to the uncertainties on the knowledge of the Dalitz distribution of 
 background events and of the $m_{ES},\, \Delta E,$ and ${\cal F}$ PDF parameters for both background and signal. Uncertainties in the efficiency variation across the Dalitz distribution are estimated. The statistical uncertainty on the  amplitude and phases of the nominal Dalitz model  also  contributes significantly  to the systematic uncertainties on the $CP$ parameters.

\begin{table}[htpb]
\begin{center}

\begin{tabular}{|c|c|c|c|c|c|c|}
\hline
   &                               \multicolumn{3}{|c|}{$B^- \to D^0 K^-$} &  \multicolumn{3}{|c|}{$B^- \to D^{*0} K^-$}  \\ \hline 
 Source                         &   \rb   & $\gamma$ & $\delta_B$ &  \rbs  & $\gamma$  & $\delta^*_B$   \\   \hline
 Combinatorial background Dalitz shape & 0.008 & $6.7^\circ$  & $3.3^\circ$    & 0.010    & $2.9^\circ$    & $5.1^\circ$        \\   \hline
 $m_{ES},\,\Delta E,\,\cal{ F}$ PDF shapes          & 0.007  & $5.4^\circ$  & $4.2^\circ$    & 0.025     & $1.8^\circ$ & $8.2^\circ$       \\   \hline
 $R$              & 0.018  & $3.1^\circ$  & $3.0^\circ$    & 0.018     & $3.1^\circ$    & $3.0^\circ$        \\   \hline
Efficiency                     & 0.004      & $3.0^\circ$  & $2.7^\circ$    & 0.005    &   $3.0^\circ$   & $2.8^\circ$   \\ \hline
 Dalitz amplitude and phase uncertainties   & 0.004  & $1.6^\circ$  & $4.7^\circ$    & 0.014     & $6.1^\circ$    & $8.8^\circ$ \\   \hline
\hline
 Total                          & 0.022  & $9.8^\circ$  & $8.3^\circ$    & 0.036     & $8.2^\circ$   & $13.8^\circ$       \\   \hline
\end{tabular}
\end{center}
\caption{Summary of the contributions to the systematic errors on \rb, $\gamma$ and $\delta_B$.}
\label{tab:syst}
\end{table}

%% file: conclusions-prl.tex
	
We report preliminary  results of the measurement of  \rb~ and of the angle $\gamma$ using the $B^-$ meson decays into $D^{0} K^-$  and $D^{*0} K^-$
 with a technique based on the Dalitz analysis of the 
$D^0 \ra K_S\pi^-\pi^+$ three-body decay. 
 From  227  million $B \bar B$ pairs   collected 
by the \babar\ detector, we reconstruct  $282 \pm 20$ $ B^- \to D^{0}K^-$, $89 \pm 11$ $B^- \to D^{*0}K^-,\, D^{*0} \to D^0\pi^0  $ and $44 \pm 8$   $B^- \to D^{\ast 0}K^-, \, D^{*0} \to D^0 \gamma$ signal events.

Values of the ratio of $b \to u$  and $b \to c$  amplitudes for the
processes $B^- \ra D^{0} K^-$ and $B^-\ra D^{*0} K^-$  at the small end of our measurements allow no determination of $\gamma$ at this statistical level.  Accounting for systematic uncertainties, we constrain these ratios to be $r_B < 0.19$ at 90\% confidence level and $r_B^\ast = 0.155 ^{+0.070}_{-0.077} \pm 0.040 \pm 0.020 $. The relative phases between  these  two amplitudes  are   $\delta_B= {114^\circ} \pm 41^\circ   \pm 8^\circ \pm 10^\circ$(${294^\circ} \pm 41^\circ \pm 8^\circ  \pm 10^\circ$) and  $\delta^*_B={303^\circ} \pm 34^\circ  \pm 14^\circ \pm 10^\circ$($ {123^\circ} \pm 34^\circ  \pm 14^\circ \pm 10^\circ$).  The first error is statistical, the second error accounts for experimental uncertainties and the third error reflects   the Dalitz model uncertainty.
  By combining the information  from  the two samples  we obtain   $\gamma= {70^\circ}  \pm 26^\circ   \pm 10^\circ \pm 10^\circ$ ($ {-110^\circ} \pm 26^\circ  \pm 10^\circ \pm 10^\circ $). For this preliminary result we have quoted confidence intervals   obtained with a Bayesian technique assuming a uniform prior  in \rb, $\gamma$ and $\delta_B$.

%% file: acknowledgements.tex
We are grateful for the 
extraordinary contributions of our \pep2\ colleagues in
achieving the excellent luminosity and machine conditions
that have made this work possible.
The success of this project also relies critically on the 
expertise and dedication of the computing organizations that 
support \babar.
The collaborating institutions wish to thank 
SLAC for its support and the kind hospitality extended to them. 
This work is supported by the
US Department of Energy
and National Science Foundation, the
Natural Sciences and Engineering Research Council (Canada),
Institute of High Energy Physics (China), the
Commissariat \`a l'Energie Atomique and
Institut National de Physique Nucl\'eaire et de Physique des Particules
(France), the
Bundesministerium f\"ur Bildung und Forschung and
Deutsche Forschungsgemeinschaft
(Germany), the
Istituto Nazionale di Fisica Nucleare (Italy),
the Foundation for Fundamental Research on Matter (The Netherlands),
the Research Council of Norway, the
Ministry of Science and Technology of the Russian Federation, and the
Particle Physics and Astronomy Research Council (United Kingdom). 
Individuals have received support from 
CONACyT (Mexico),
the A. P. Sloan Foundation, 
the Research Corporation,
and the Alexander von Humboldt Foundation.

%% file: conf_sum2004.bbl
\begin{thebibliography}{99}

\bibitem{ref:CP}          \babar\ Collaboration, B. Aubert {\it et al.}, Phys. Rev. Lett. {\bf 89}, 201802 (2002);~
                          Belle Collaboration, K. Abe {\it et al.} Phys. Rev. {\bf D66}, 071102 (2002).
\bibitem{ref:CKM}
N.~Cabibbo, \jprl {\bf 10}, 531 (1963);
M.~Kobayashi and T.~Maskawa, \progtp {\bf 49}, 652 (1973).
\bibitem{ref:Wolf} L.Wolfenstein,  Phys. Rev. Lett. {\bf 51} 1945 (1983).

\bibitem{chargeconj} Reference to the 
charge-conjugate state is implied here and throughout 
the text unless otherwise stated.

\bibitem{ref:GLW}     M. Gronau and D. London,  Phys. Lett. {\bf B253}, 483 (1991);~
                          M. Gronau and D. Wyler,  Phys. Lett.{\bf B265}, 172 (1991).
\bibitem{ref:ADS}                    I. Dunietz,  Phys. Lett. {\bf B270}, 75 (1991);~
                          I. Dunietz,  Z. Phys. {\bf C56}, 129 (1992);~
                          D. Atwood, G. Eilam, M. Gronau and A. Soni,  Phys. Lett. {\bf B341}, 372 (1995);~
                          D. Atwood, I. Dunietz and A. Soni,  Phys. Rev. Lett. {\bf 78}, 3257 (1997).

\bibitem{ref:DKDalitz}    A. Giri, Yu. Grossman, A. Soffer and J. Zupan, Phys. Rev. {\bf D68}, 054018 (2003).

\bibitem{Poluektov:2004mf}
A.~Poluektov {\it et al.}  [Belle Collaboration],
arXiv:hep-ex/0406067.
\bibitem{ref:bondar}
A.~Bondar and T.~Gershon,
arXiv:hep-ph/0409281.

\bibitem{ref:detector}    \babar\ Collaboration, B. Aubert {\it et al.}, Nucl. Instr. Meth. {\bf A 479}, 1 (2002).  
\bibitem{ref:pdg2004}     Particle Data Group, S. Eidelman {\it et al.}, Phys. Lett. {\bf B592}, 1 (2004)   



\bibitem{ref:cleomodel}   S. Kopp et al. (CLEO Coll.)  Phys. Rev.  {\bf D63} 092001 (2001);~
                          H. Muramatsu et al. (CLEO Coll.)  Phys. Rev. Lett.  {\bf 89} 251802 (2002);~
                          Erratum-ibid: {\bf 90} 059901 (2003).
%

\bibitem{ref:gounarissakurai} G.J. Gounaris and J.J. Sakurai,  Phys. Rev. Lett.  {\bf 21} 244, (1968);

\bibitem{ref:argus} 
\begin{math}
\frac{dN}{dm_{ES}} = N\cdot \mes \cdot \sqrt{1-x^2} \cdot \exp \left(- \xi \cdot (1 - x^2)\right)
\end{math}
where $x = 2\mes/\sqrt{s}$ and the parameter $\xi$ is determined from a fit.
 ARGUS Collaboration, H.~Albrecht {\em et al.}, \zpc{48}, 543 (1990).


\bibitem{dspaper}  \rb~ in $B^- \to  D^{(*)0} \pi^-$ can be roughly estimated as $\frac{|V^*_{ub}V_{cd}|}{|V^*_{cb}V_{ud}|}  \cdot \frac{1}{3}$$\sim$$~0.007$ where $\frac{1}{3}$ accounts for the color suppression.

%

\end{thebibliography}
